\documentclass{article}

\PassOptionsToPackage{dvipsnames}{xcolor}

\usepackage[final,nonatbib]{neurips_2026}

\usepackage[numbers]{natbib}
\usepackage{hyperref}
\usepackage{url}
\usepackage{booktabs}
\usepackage{nicefrac}
\usepackage{multirow}
\usepackage{subcaption}
\usepackage[breakable]{tcolorbox}

\title{QCalEval: Benchmarking Vision-Language Models for Quantum Calibration Plot Understanding}

\author{\parbox{\textwidth}{\centering
  \mbox{Shuxiang Cao$^{1}$\thanks{\texttt{shuxiangc@nvidia.com}}},
  \mbox{Zijian Zhang$^{1,2,13}$},
  \mbox{Abhishek Agarwal$^{6}$},
  \mbox{Grace Bratrud$^{10,9}$},
  \mbox{Niyaz R.\ Beysengulov$^{11}$},
  \mbox{Daniel C.\ Cole$^{7}$},
  \mbox{Alejandro G\'omez Frieiro$^{3}$},
  \mbox{Elena O.\ Glen$^{11}$},
  \mbox{Hao Hsu$^{3}$},
  \mbox{Gang Huang$^{4}$},
  \mbox{Raymond Jow$^{5}$},
  \mbox{Greshma Shaji$^{3}$},
  \mbox{Tom Lubowe$^{1}$},
  \mbox{Ligeng Zhu$^{1}$},
  \mbox{Luis Mantilla Calder\'on$^{1,2,13}$},
  \mbox{Nicola Pancotti$^{1}$},
  \mbox{Joel Pendleton$^{5}$},
  \mbox{Brandon Severin$^{5}$},
  \mbox{Charles Etienne Staub$^{8}$},
  \mbox{Sara Sussman$^{9}$},
  \mbox{Antti Veps\"al\"ainen$^{3}$},
  \mbox{Neel Rajeshbhai Vora$^{4}$},
  \mbox{Yilun Xu$^{4}$},
  \mbox{Varinia Bernales$^{2}$},
  \mbox{Daniel Bowring$^{9}$},
  \mbox{Elica Kyoseva$^{1}$},
  \mbox{Ivan Rungger$^{6,12}$},
  \mbox{Giulia Semeghini$^{8}$},
  \mbox{Sam Stanwyck$^{1}$},
  \mbox{Timothy Costa$^{1}$},
  \mbox{Al\'an Aspuru-Guzik$^{1,2,13}$},
  \mbox{Krysta Svore$^{1}$}
  \\[0.5em]
  {\small
  \mbox{$^{1}$NVIDIA},
  \mbox{$^{2}$University of Toronto},
  \mbox{$^{3}$IQM Quantum Computers},
  \mbox{$^{4}$Lawrence Berkeley National Laboratory},
  \mbox{$^{5}$Conductor Quantum},
  \mbox{$^{6}$National Physical Laboratory},
  \mbox{$^{7}$Infleqtion},
  \mbox{$^{8}$Harvard University},
  \mbox{$^{9}$Fermi National Accelerator Laboratory},
  \mbox{$^{10}$Northwestern University},
  \mbox{$^{11}$EeroQ Corporation},
  \mbox{$^{12}$Royal Holloway University of London},
  \mbox{$^{13}$Vector Institute for Artificial Intelligence}}
}}

\begin{document}

\maketitle

\begin{abstract}
Quantum computing calibration depends on interpreting experimental data, and calibration plots provide the most universal human-readable representation for this task, yet no systematic evaluation exists of how well vision-language models (VLMs) interpret them.
We introduce \textbf{QCalEval}, the first VLM benchmark for quantum calibration plots: 243 samples across 87 scenario types from 22 experiment families, spanning superconducting qubits and neutral atoms, evaluated on six question types in both zero-shot and in-context learning settings.
The best general-purpose zero-shot model reaches a mean score of 72.3, and many open-weight models degrade under multi-image in-context learning, whereas frontier closed models improve substantially.
A supervised fine-tuning ablation at the 9-billion-parameter scale shows that SFT improves zero-shot performance but cannot close the multimodal in-context learning gap.
As a reference case study, we release \textbf{NVIDIA Ising Calibration 1}, an open-weight model based on Qwen3.5-35B-A3B that reaches 74.7 zero-shot average score.

\end{abstract}

\section{Introduction}
\label{sec:intro}

Quantum computing systems require continuous calibration to characterize and maintain their operating parameters, as quantum states are sensitive to environmental perturbations.
Key calibration targets, including transition frequencies, pulse amplitudes, readout settings, trapping conditions, and couplings, vary by platform and drift over time due to environmental fluctuations and hardware instabilities~\cite{wittler2021integrated,werninghaus2021highspeed,agarwal2025fluctuations}.
As systems scale to hundreds of qubits and beyond, the calibration burden grows combinatorially: each qubit requires dozens of characterization experiments, and the results of one calibration step can invalidate others, creating complex dependency chains~\cite{pasquale2023qibocal,kanazawa2023qiskit}. Similarly, results from holistic benchmarking of quantum computers can produce large amounts of data, making it challenging to interpret dependencies across metrics and to analyze the effects of different sources of error~\cite{lall2025reviewcollectionmetricsbenchmarks}. 
Automating calibration analysis, from interpreting results to reasoning about next steps, is therefore essential to scale calibration and tuning workflows reliably, yet common approaches often still rely on manual or semi-manual calibration tasks.
The standard approach to analyzing calibration data is \emph{parametric model fitting}, where an assumed functional form (e.g., a decaying sinusoid for Rabi oscillations) is fit to the measured data.
Practitioners assess reliability using goodness-of-fit measures such as $R^2$ or $\chi^2$, residual structure, parameter uncertainty, and platform-specific heuristics before feeding extracted parameters back into the system.

\begin{figure}[t]
    \centering
    \includegraphics[width=\textwidth]{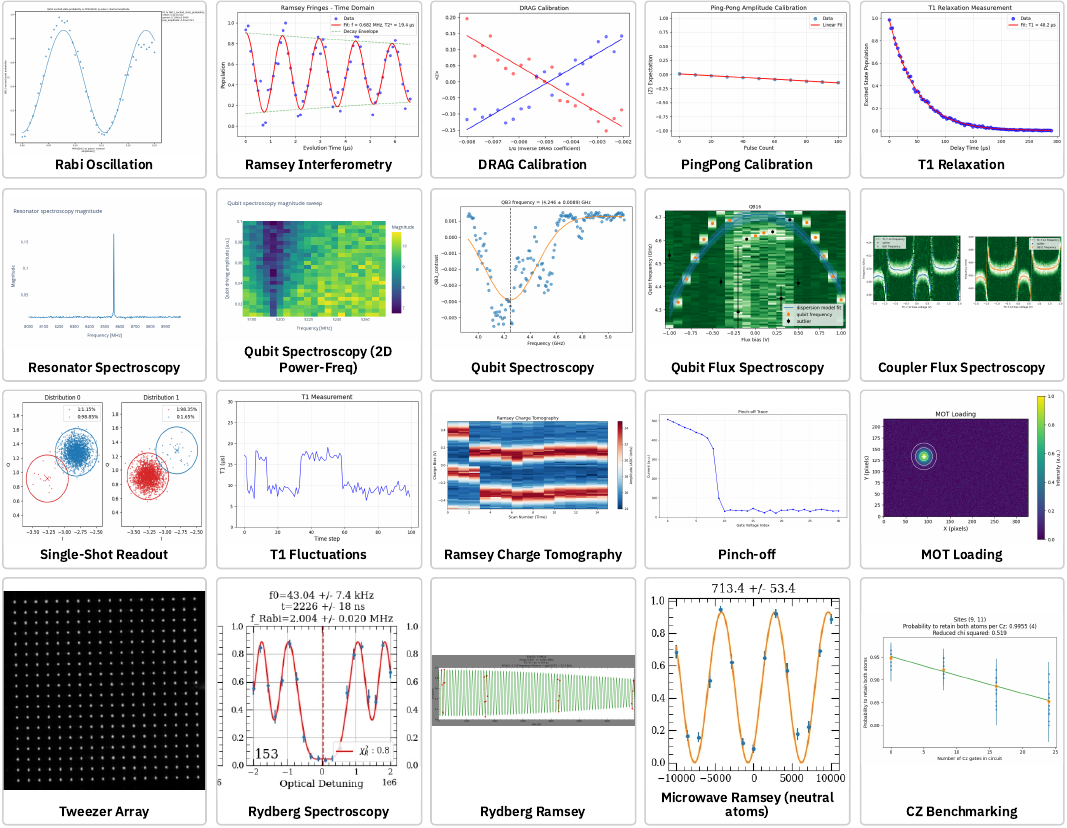}
    \caption{Representative calibration plots from QCalEval. The benchmark is visually heterogeneous: it includes 1D line traces with oscillations and decays, 2D spectroscopy maps with ridges and hotspots, scatter or histogram-style readout diagnostics, and image-like measurements of spatial structure. Unlike natural-image benchmarks, the key information is carried by scientific geometry rather than object identity: peak locations, fringe spacing, contrast, linewidth, clustering, and the presence or absence of fitted structure determine whether an experiment is reliable or unreliable.}
    \label{fig:experiments}
\end{figure}

Recent work has moved toward \emph{agentic} calibration in response to the combinatorial scaling of qubit calibration tasks and need to move beyond manual calibration, where AI agents autonomously orchestrate multi-step calibration workflows.
Some of us, in Cao et al.~\cite{cao2025kagents}, introduced \textit{k-agents}, a system where LLM-based agents decide which experiments to run, interpret results, and adapt the calibration strategy in real time.
A key bottleneck for agentic systems such as \textit{k-agents} is \emph{interpreting calibration plots}: the agent must examine an experimental result (typically a plot) to determine whether the experiment succeeded, what went wrong if it failed, and what to do next.

Calibration plots are the \emph{universal human-readable representation} of calibration results. Regardless of hardware platform, software stack, or underlying raw measurement format, any experiment must ultimately be rendered in visual form.
This motivates the use of vision-language models (VLMs) as the ``eyes'' of calibration agents, since a single vision-based model can interpret any calibration experiment without platform-specific integration.
This approach is \emph{complementary} to parametric fitting, which assumes a correct model and fails silently when violated~\cite{agarwal2024nonmarkovian}, whereas a VLM trained on diverse failure modes can catch failures that no single parametric model anticipates.

Despite this potential, no systematic evaluation exists for how well VLMs handle quantum calibration plots.
VLMs have demonstrated remarkable capabilities in natural image understanding~\cite{bai2023qwenvl,openai2023gpt4v,liu2023llava}, and multimodal in-context learning (MM-ICL) enables adaptation to new tasks through demonstration examples~\cite{alayrac2022flamingo}.
However, interpreting calibration plots requires both identifying domain-specific geometric features (oscillation frequencies, decay rates, peak positions) and mapping them to operational statuses, demanding precise visual perception alongside expert domain knowledge.
Whether current VLMs can reliably interpret these plots, and whether in-context demonstrations help or hinder, remain open questions.

\begin{figure}[t]
    \centering
    \includegraphics[width=\textwidth]{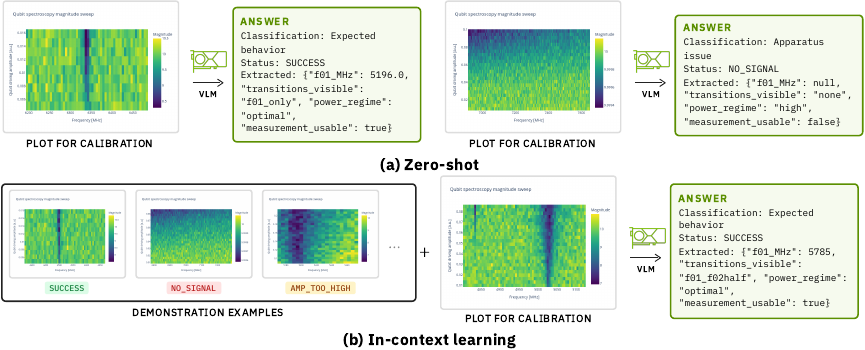}
    \caption{Task illustration with examples from the qubit spectroscopy experiment. Models are evaluated under both zero-shot (single plot, no demonstrations) and in-context learning (demonstration examples provided) settings. Each evaluation is an independent conversation with no shared context.}
    \label{fig:task}
\end{figure}

We introduce \textbf{QCalEval}, the first comprehensive benchmark for VLMs on quantum calibration plots, evaluating models under both zero-shot (no examples provided) and in-context learning (ICL; with demonstration examples) conditions (Figure~\ref{fig:task}).
In zero-shot mode, the model receives a single calibration plot and must answer without prior examples; in ICL mode, labeled examples from the same experiment family are provided before the query plot.
Our benchmark of 18 VLMs across six question types establishes the first baseline score for this domain: even the best general-purpose zero-shot model achieves a mean score of 72.3. Under in-context learning evaluation, frontier closed models and Gemma 4 improve substantially (up to +29 scores on calibration diagnosis), while many open-weight models degrade with multi-image prompts.
A systematic supervised fine-tuning (SFT) ablation (5 recipes at the 9-billion-parameter (9B) scale using Qwen3.5), where training data is formatted either as zero-shot queries (single plot) or ICL queries (with demonstration plots), shows that SFT improves zero-shot performance but cannot close the multimodal in-context learning gap.

\vspace{0.5em}

\noindent\textbf{Contributions.}
\begin{enumerate}

    \item \textbf{QCalEval Benchmark}: The first comprehensive benchmark for VLMs on quantum calibration plots, comprising 243 samples across 87 scenario types from 22 experiment families spanning superconducting qubits and neutral atoms, evaluated on six question types under both zero-shot and in-context learning settings.
    \item \textbf{Zero-Shot Baseline}: We evaluate 18 VLMs and establish the first score baseline for understanding quantum calibration plots, finding that even the best general-purpose model reaches a mean score of 72.3, revealing this domain as a significant challenge for current VLMs.
    \item \textbf{MM-ICL Gap}: When provided with labeled visual examples of each scenario type before the query plot, frontier closed models and Gemma improve strongly, while many open-weight models perform \emph{worse} than without any examples, exposing a clear multimodal in-context learning gap.
    \item \textbf{SFT Ablation}: We systematically evaluate 5 SFT recipes at 9B scale with Qwen3.5, finding that the strongest sequential curriculum in the ablation study is ICL$\rightarrow$zero-shot, while no configuration fixes free-text scientific reasoning (Q3) under in-context learning. 
    \item \textbf{NVIDIA Ising Calibration 1}: Guided by the ablation findings, we release an open-weight 35B mixture-of-experts (MoE) model trained with the strongest sequential curriculum identified in that study (two-phase SFT: ICL then zero-shot) as a reference case study.
\end{enumerate}

\section{Related Work}
\label{sec:related}

\paragraph{Vision-Language Models and Multimodal In-Context Learning.}
Modern VLMs combine pretrained vision encoders with large language backbones via alignment objectives such as CLIP~\cite{radford2021clip} or modular recipes such as BLIP-2~\cite{li2023blip2}, with instruction tuning converting them into visual assistants~\cite{dai2023instructblip,liu2023llava}.
Flamingo~\cite{alayrac2022flamingo} and OpenFlamingo~\cite{awadalla2023openflamingo} introduced interleaved image-text sequences for few-shot multimodal adaptation, but strong zero-shot performance does not guarantee robust multimodal in-context learning (MM-ICL): VL-ICL Bench~\cite{zong2024vlicl} and recent analyses~\cite{doveh2024icl,jiang2024manyshot} show that the effectiveness of in-context learning demonstrations is fragile and highly sensitive to prompt construction.

\paragraph{Chart Understanding and Scientific Figures.}
Chart reasoning benchmarks evolved from FigureQA~\cite{kahou2017figureqa} and DVQA~\cite{kafle2018dvqa} to PlotQA~\cite{methani2020plotqa} and ChartQA~\cite{masry2022chartqa}, revealing that chart reasoning depends on OCR, numerical grounding, and structural relations~\cite{hoque2022cqasurvey}.
Chart-specific models include ChartOCR~\cite{luo2021chartocr}, DePlot~\cite{liu2023deplot}, UniChart~\cite{kavehzadeh2023unichart}, ChartLlama~\cite{han2023chartllama}, ChartInstruct~\cite{masry2024chartinstruct}, and ChartGemma~\cite{masry2024chartgemma}, though evaluations show VLMs remain error-prone on scientific figures~\cite{islam2024charteval,roberts2024scifibench}.
Related scientific-figure resources, such as SciCap~\cite{hsu2021scicap} and Multimodal ArXiv~\cite{li2024arxivcap}, broaden this line beyond standard charts but still do not target operational diagnosis over calibration plots.
These benchmarks focus on general charts rather than expert-oriented diagnosis over quantum calibration plots.
Notably, chart-specific pipelines such as DePlot~\cite{liu2023deplot} (plot-to-table translation followed by LLM reasoning), UniChart~\cite{kavehzadeh2023unichart} (chart-specific pretraining), and ChartGemma~\cite{masry2024chartgemma} (visual instruction tuning for charts) are designed for data extraction and QA on standard chart types (bar, line, pie).
They are not evaluated in our benchmark because quantum calibration requires \emph{domain-specific diagnostic reasoning}, such as determining whether an oscillation frequency is too fast, whether a fit deviation indicates a model failure, or whether a measurement window is sufficient, rather than reading values from axes or answering factual questions about plotted data.
Extending these chart-specific approaches to scientific calibration diagnosis is an interesting direction for future work.

\paragraph{Quantum Calibration and Automation.}
Quantum calibration is an iterative process where practitioners inspect oscillations, decays, and spectroscopy scans to tune device parameters~\cite{rol2017restless,werninghaus2021highspeed}, with integrated software stacks for control and analysis~\cite{wittler2021integrated,kanazawa2023qiskit,pasquale2023qibocal}.
Recent work has begun applying LLMs directly to quantum experiments, including agent-based calibration workflows and instruction-to-experiment translation for superconducting qubit experiments~\cite{cao2025kagents,li2026llmassistedqubit}, while broader reviews identify calibration and recovery as core automation bottlenecks for scalable quantum computing~\cite{alexeev2025aiquantum}.
Our benchmark evaluates whether VLMs can interpret calibration plots as actionable artifacts for quantum hardware workflows.

\paragraph{Scientific Instrument Agents and Autonomous Laboratories.}
Across microscopy, beamlines, and autonomous chemistry, LLMs are increasingly used as tool-augmented scientific assistants that translate natural-language goals into scripts, retrieve documentation, orchestrate device APIs, and operate within constrained workflow engines~\cite{skreta2023clairify,boiko2023coscientist,darvish2024organa,ruan2024synthesis,mandal2025afmbench,xie2025instrumentcontrol,vriza2026calms}.
These systems motivate our focus on execution-adjacent scientific workflows, but they do not benchmark fine-grained visual diagnosis of quantum calibration plots.

\section{QCalEval Benchmark}
\label{sec:benchmark}

QCalEval evaluates VLM capabilities on quantum calibration plots through six question types, assessed under both \emph{zero-shot} (no demonstrations) and \emph{in-context learning} (with demonstrations) settings.

\paragraph{Benchmark Scope.}
QCalEval mainly covers \textbf{superconducting qubits} and \textbf{neutral atoms}, along with emerging platforms, including both shared calibration routines and platform-specific diagnostics.
The 22 experiment families span a wide range of visual formats---1D line traces, 2D spectroscopy maps, histograms, and image-like measurements---as illustrated by the representative examples in Figure~\ref{fig:experiments}.
This mix tests whether VLMs can generalize across familiar calibration patterns as well as domain-specific artifacts and failure modes.
The benchmark includes both simulated and real-hardware data provided by multiple partners~\cite{abdurakhimov2024iqm,bratrud2025chargenoise}; Table~\ref{tab:benchmark} summarizes the benchmark scale and data-source split, and the full list of experiment families is provided in the appendix.
Each experiment family defines multiple scenario types (typically two to seven), such as different success and failure modes; each scenario type contains one or more benchmark samples, where each sample comprises one or more calibration plot images, a scenario type label, and ground-truth answers for all six question types.

\paragraph{Task Taxonomy and Motivation.}
We define six question types because quantum calibration assistance is not a single visual question-answering problem, but a pipeline from accurately perceiving a plot to making an operational calibration decision.
Each question isolates a different failure mode that would be hidden by a single aggregate score:
\begin{itemize}
    \item \textbf{Q1} (\emph{technical description}): a structured JSON description of the plot type, axes, and salient visual features, isolating visual grounding from domain reasoning.
    \item \textbf{Q2} (\emph{experimental conclusion}): a coarse 4-way outcome classification (\texttt{Expected behavior}, \texttt{Suboptimal parameters}, \texttt{Anomalous behavior}, \texttt{Apparatus issue}), testing whether the model can map visual evidence to an experimental interpretation.
    \item \textbf{Q3} (\emph{experimental significance}): experiment-specific scientific analysis of what the observed pattern implies, whether the sweep window or resolution is sufficient, and what next calibration step should follow.
    \item \textbf{Q4} (\emph{fit reliability}): whether a visible fit, if present, is trustworthy for downstream use, forcing a decision over \texttt{Reliable}, \texttt{Unreliable}, or \texttt{No fit}.
    \item \textbf{Q5} (\emph{parameter extraction}): machine-readable extraction of family-specific physical parameters in structured JSON.
    \item \textbf{Q6} (\emph{calibration diagnosis}): the most operational task, assigning a family-specific status code (e.g., \texttt{SUCCESS}, \texttt{NO\_SIGNAL}) and, when needed, providing a corrective range or suggested action.
\end{itemize}
This decomposition mirrors how experimentalists actually use calibration plots: first identify what is shown, then judge what broad outcome it represents, reason about its scientific meaning, assess whether quantitative readout is justified, extract any usable parameters, and finally decide what action to take.
We intentionally include both Q2 and Q6 because they operate at different granularities: Q2 is a shared coarse outcome label across all experiment families, while Q6 is a family-specific actionable diagnosis.

\begin{figure}[t]
    \centering
    \includegraphics[width=\textwidth]{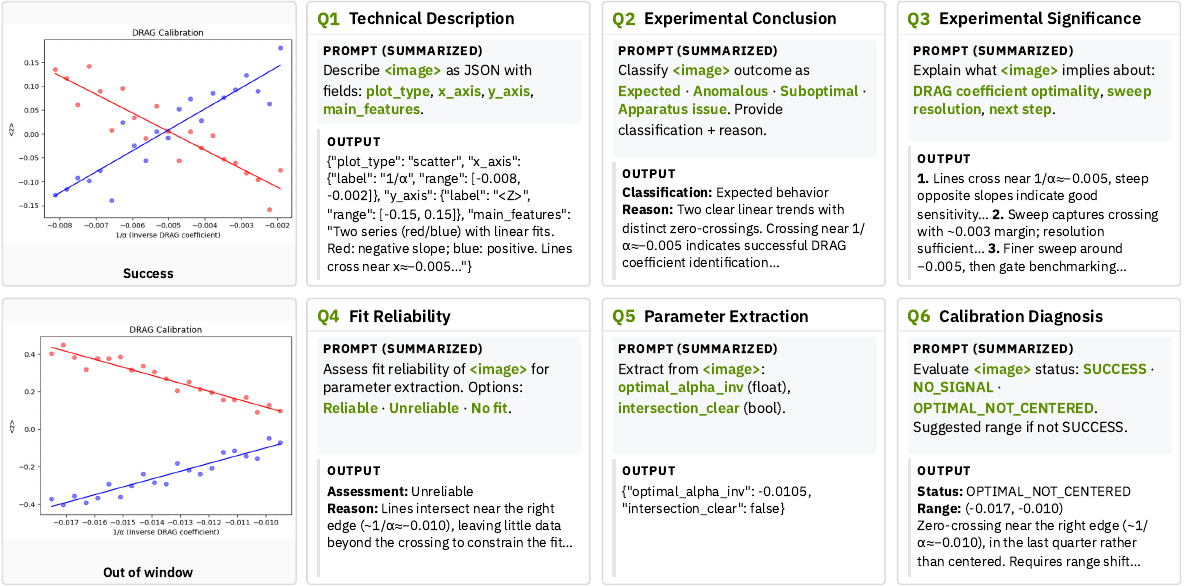}
    \caption{The six question types in QCalEval, illustrated on a DRAG calibration example. Q1 uses a universal prompt across all families; Q2--Q6 prompts are customized per experiment family with family-specific background context, failure mode definitions, and parameter extraction schemas.}
    \label{fig:questions}
\end{figure}

\paragraph{Zero-Shot Evaluation.}
Each sample is evaluated on all six question types without any demonstration examples (Figure~\ref{fig:questions}): the model receives only the query plot and textual background.
For Q2--Q6, prompts include textual domain knowledge, provided as part of the released benchmark for all 22 experiment families: each background describes what the experiment measures, what a successful result looks like, and the set of possible outcome or status labels.
This establishes a baseline for how well VLMs understand quantum calibration plots in the absence of visual examples, using only textual context.

\paragraph{In-Context Learning Evaluation.}
In-context learning provides demonstration examples (previously collected calibration plots with expert-assigned labels) of each scenario type, directly showing the model what each failure mode looks like.
These demonstrations are drawn from other samples within the same experiment family in the benchmark, mirroring how practitioners in practice would reference past calibration runs with known outcomes to interpret a new result.
Three of the six question types (Q3, Q5, Q6) are additionally evaluated with demonstration examples, each posed as an \emph{independent conversation} with no shared context:
\begin{enumerate}
    \item \textbf{$N$-way In-Context Analysis (Q3).} The model receives one demonstration per scenario type in the experiment family, each pairing a scenario plot with an expert analysis describing the observed behavior, its diagnostic implications, and recommended next steps. Given a new query plot, the model must produce a similarly structured analysis.

    \item \textbf{1-Shot Parameter Extraction (Q5).} The model receives a single worked example showing a plot and its extracted parameters as a structured JSON object (e.g., \texttt{\{``optimal\_alpha\_inv'': -0.005, ``intersection\_clear'': true\}}). Given a new query plot, the model must extract analogous parameters.

    \item \textbf{$N$-way In-Context Classification (Q6).} The model receives one labeled example per scenario type in the experiment family (e.g., \texttt{SUCCESS}, \texttt{NO\_SIGNAL}), and must classify a new query plot into one of these scenarios.
\end{enumerate}
Scenario types with only one sample are excluded from in-context learning evaluation because they cannot provide support examples without reusing the query itself.
We restrict in-context learning evaluation to Q3, Q5, and Q6 because these are the tasks where demonstrations can plausibly transfer structured reasoning patterns, output schemas, or family-specific label vocabularies.
By contrast, Q1, Q2, and Q4 already have tightly specified zero-shot output spaces and are less informative probes of MM-ICL.
Comparing zero-shot and in-context learning scores on Q3, Q5, and Q6 directly measures whether models can exploit in-context demonstrations.

\begin{table}[t]
    \centering
    \caption{QCalEval benchmark statistics.}
    \label{tab:benchmark}
    \begin{tabular}{lr|lr}
        \toprule
        \multicolumn{2}{c|}{\textbf{Dataset Statistics}} & \multicolumn{2}{c}{\textbf{Benchmark Evaluation}} \\
        \midrule
        Benchmark samples & 243 & Zero-shot & 243 samples $\times$ 6 Qs \\
        Scenario types & 87 & In-context learning & 236 samples $\times$ 3 Qs \\
        Experiment families & 22 & & \\
        Unique images & 309 & & \\
        \bottomrule
    \end{tabular}
\end{table}

\section{Results and Analysis}
\label{sec:evaluation}

We evaluate 18 VLMs spanning frontier closed-source models (GPT-5.4~\cite{openai_gpt54_2025}, Gemini~3.1~\cite{google2026gemini31pro}, Claude~4.6~\cite{anthropic_claude_opus_46_2025}), open-weight models (Qwen3.5~\cite{qwen3.5}, Gemma~4~\cite{googledeepminds2026gemma4}, InternVL3~\cite{zhu2025internvl3}, Kimi-VL~\cite{team2025kimi}, MiniCPM-o~\cite{yao2024minicpm}), and one domain-tuned case-study model \textbf{NVIDIA Ising Calibration 1} (Ising-Cal-1) on all six QCalEval question types.
Q2, Q4, Q5, and Q6 are scored programmatically; Q1 and Q3 are scored by two independent LLM judges (GPT-5.4 and Gemini~3.1~Pro) and averaged to reduce single-judge bias (details in Appendix~\ref{app:protocol}).
We report zero-shot and in-context learning (MM-ICL) results.
We additionally conduct an $N$-way scaling study on Q6, varying the number of demonstration examples from 0 to 5 shots across four experiment families with five scenario types each, to probe whether the MM-ICL gap is caused by image overload.
Table~\ref{tab:summary} summarizes the best-performing model for each question type across all evaluation settings; full per-model results appear in Tables~\ref{tab:zeroshot} and~\ref{tab:icl}.

\begin{table}[t]
    \centering
    \caption{Best score per question type across all evaluation settings. Q1, Q2, and Q4 are evaluated under zero-shot only; Q3, Q5, and Q6 are evaluated under both zero-shot and ICL. $^\dagger$Domain-tuned case study (Section~\ref{sec:model}).}
    \label{tab:summary}
    \small
    \begin{tabular}{llc}
        \toprule
        \textbf{Task} & \textbf{Best Model} & \textbf{Score} \\
        \midrule
        Q1 (Description) & GPT-5.4 & 90.9 \\
        Q2 (Conclusion) & Ising-Cal-1-35B$^\dagger$ & 67.1 \\
        Q3 (Significance) & Claude Opus 4.6 (ICL) & 84.7 \\
        Q4 (Fit reliability) & Ising-Cal-1-35B$^\dagger$ & 90.5 \\
        Q5 (Param.\ extraction) & Gemini-3.1-Pro (ICL) & 84.5 \\
        Q6 (Diagnosis) & Gemini-3.1-Pro (ICL) & 89.8 \\
        \bottomrule
    \end{tabular}
\end{table}

\subsection{Zero-Shot Performance}

\begin{table}[t]
    \centering
    \caption{Zero-shot scores across six evaluation axes. Models grouped by access type. \textbf{Bold} = best per column. $^\dagger$Domain-tuned case study (Section~\ref{sec:model}).}
    \label{tab:zeroshot}
    \small
    \begin{tabular}{llccccccr}
        \toprule
        & \textbf{Model} & \textbf{Q1} & \textbf{Q2} & \textbf{Q3} & \textbf{Q4} & \textbf{Q5} & \textbf{Q6} & \textbf{Mean} \\
        \midrule
        \multirow{7}{*}{\rotatebox{90}{\scriptsize Closed}}
        & Gemini-3.1-Pro & 88.5 & 57.2 & 61.1 & 84.4 & \textbf{71.5} & 71.2 & 72.3 \\
        & Claude Opus 4.6 & 90.8 & 49.0 & \textbf{65.5} & 76.1 & 64.7 & 60.5 & 67.8 \\
        & Gemini-3.1-Flash-Lite & 89.2 & 53.5 & 59.4 & 82.7 & 63.8 & 60.9 & 68.2 \\
        & Claude Sonnet 4.6 & 89.7 & 48.6 & 63.4 & 76.5 & 60.4 & 60.1 & 66.5 \\
        & GPT-5.4 & \textbf{90.9} & 52.7 & 63.7 & 54.7 & 64.3 & 61.3 & 64.6 \\
        & GPT-5.4-Mini & 90.3 & 39.5 & 48.3 & 42.0 & 62.6 & 51.4 & 55.7 \\
        & Claude Haiku 4.5 & 83.4 & 36.6 & 40.8 & 48.6 & 51.0 & 42.8 & 50.5 \\
        \midrule
        \multirow{11}{*}{\rotatebox{90}{\scriptsize Open}}
        & Gemma-4-31B-IT & 85.6 & 54.3 & 59.8 & 82.7 & 68.3 & 62.1 & 68.8 \\
        & Qwen3.5-397B-A17B & 88.1 & 42.8 & 52.0 & 50.6 & 62.5 & 55.6 & 58.6 \\
        & Qwen3.5-27B & 87.0 & 45.7 & 48.3 & 56.4 & 58.7 & 55.1 & 58.5 \\
        & Qwen3.5-122B-A10B & 86.6 & 44.0 & 49.0 & 50.2 & 61.2 & 51.9 & 57.1 \\
        & Qwen3.5-35B-A3B & 86.8 & 39.9 & 45.7 & 52.7 & 57.8 & 50.6 & 55.5 \\
        & Qwen3.5-9B & 81.5 & 37.9 & 39.5 & 49.8 & 57.1 & 52.3 & 53.0 \\
        & InternVL3-78B & 76.3 & 37.0 & 34.1 & 42.8 & 52.9 & 45.7 & 48.2 \\
        & MiniCPM-o-4.5 & 76.7 & 31.7 & 29.8 & 32.5 & 47.9 & 48.1 & 44.5 \\
        & InternVL3-38B & 79.2 & 34.6 & 27.6 & 33.7 & 49.2 & 40.3 & 44.1 \\
        & Kimi-VL-A3B & 65.0 & 34.6 & 22.1 & 35.0 & 38.9 & 37.4 & 38.9 \\
        \cmidrule{2-9}
        & \textbf{Ising-Cal-1-35B}$^\dagger$ & 87.8 & \textbf{67.1} & 64.7 & \textbf{90.5} & 62.5 & \textbf{75.3} & \textbf{74.7} \\
        \bottomrule
    \end{tabular}
\end{table}

\paragraph{Finding 1: Models detect visual features well but lack domain knowledge to interpret them.}
The best base model (Gemini-3.1-Pro) achieves a mean score of 72.3; the domain-tuned Ising-Cal-1 reaches 74.7.
Gemini-3.1-Pro leads on parameter extraction (Q5: 71.5) while Ising-Cal-1 leads on outcome classification (Q2: 67.1), fit assessment (Q4: 90.5), and calibration diagnosis (Q6: 75.3), reflecting the benefit of domain tuning on tasks that require expert knowledge.
Interpreting calibration plots requires two capabilities: detecting relevant visual features and mapping them to domain-specific operational outcomes.
Models perform well on feature detection (Q1: 65--91\%), but performance drops sharply on tasks requiring domain knowledge: outcome classification (Q2: 32--67\%) and calibration diagnosis (Q6: 37--75\%).
The primary failure mode is optimistic bias: across all models, 60.7\% of ``Suboptimal parameters'' cases are classified as ``Expected behavior'' --- models see the features but default to success without the domain knowledge to interpret them.
Difficulty scales with domain knowledge required: families where failure is visually obvious, such as resonator spectroscopy (100\%), are solved by all models, while families where failures resemble successful experiments, such as single-shot readout and Ramsey T2*, require domain expertise to identify (Appendix~\ref{app:perfamily}).

\paragraph{Finding 2: Fit assessment exposes a visual judgment gap.}
Fit assessment (Q4: 33--84\%) does not require domain expertise but tests whether models can judge if a fitted curve matches the underlying data.
Two failure modes emerge: \emph{false confidence}, where models predict ``Reliable'' for unreliable fits because the curve visually looks plausible despite poor fit quality (ranging from 6\% for Gemini-3.1-Flash-Lite to 74\% for InternVL3-38B); and \emph{no-fit blindness}, where models label raw data without fitted curves as ``Unreliable'' rather than ``No fit'' (ranging from 0\% for Ising-Cal-1, Gemma-4-31B, and Gemini-3.1-Pro to 91\% for InternVL3-38B).
Models that learn to output ``No fit'' (matching the 31.7\% ground-truth rate) perform better than those that almost never predict it.

\subsection{In-Context Learning: Can Demonstrations Help?}

\begin{table}[t]
    \centering
    \caption{In-context learning (MM-ICL) scores. Q3 uses $N$-way analysis demos, Q5 uses 1-shot extraction demos, Q6 uses $N$-way classification demos. $\Delta$ = change from zero-shot. \textcolor{teal}{green} = improvement, \textcolor{red}{red} = degradation.}
    \label{tab:icl}
    \small
    \begin{tabular}{llcrcrcrc}
        \toprule
        & \textbf{Model} & \textbf{Q3} & \textbf{$\Delta$} & \textbf{Q5} & \textbf{$\Delta$} & \textbf{Q6} & \textbf{$\Delta$} & \textbf{Mean} \\
        \midrule
        & \textit{Best zero-shot (base)} & \textit{65.5} & & \textit{71.5} & & \textit{71.2} & & \\
        \midrule
        \multirow{7}{*}{\rotatebox{90}{\scriptsize Closed}}
        & Gemini-3.1-Pro & 81.3 & \color{teal}{+20.2} & \textbf{84.5} & \color{teal}{+13.0} & \textbf{89.8} & \color{teal}{+18.6} & \textbf{85.2} \\
        & Claude Opus 4.6 & \textbf{84.7} & \color{teal}{+19.2} & 81.3 & \color{teal}{+16.6} & 89.4 & \color{teal}{+28.9} & 85.1 \\
        & Claude Sonnet 4.6 & 77.8 & \color{teal}{+14.4} & 71.9 & \color{teal}{+11.5} & 78.0 & \color{teal}{+17.9} & 75.9 \\
        & Gemini-3.1-Flash-Lite & 78.5 & \color{teal}{+19.1} & 73.6 & \color{teal}{+9.8} & 82.2 & \color{teal}{+21.3} & 78.1 \\
        & GPT-5.4 & 81.0 & \color{teal}{+17.3} & 72.9 & \color{teal}{+8.6} & 81.4 & \color{teal}{+20.1} & 78.4 \\
        & GPT-5.4-Mini & 58.8 & \color{teal}{+10.5} & 72.7 & \color{teal}{+10.1} & 66.9 & \color{teal}{+15.5} & 66.1 \\
        & Claude Haiku 4.5 & 66.1 & \color{teal}{+25.3} & 58.7 & \color{teal}{+7.7} & 73.1 & \color{teal}{+30.3} & 66.0 \\
        \midrule
        \multirow{10}{*}{\rotatebox{90}{\scriptsize Open}}
        & Gemma-4-31B-IT & 80.6 & \color{teal}{+20.8} & 76.9 & \color{teal}{+8.6} & 86.0 & \color{teal}{+23.9} & 81.2 \\
        & InternVL3-38B & 56.2 & \color{teal}{+28.6} & 59.5 & \color{teal}{+10.3} & 55.1 & \color{teal}{+14.8} & 56.9 \\
        & Qwen3.5-27B & 41.8 & \color{red}{-6.5} & 71.5 & \color{teal}{+12.8} & 45.8 & \color{red}{-9.3} & 53.0 \\
        & InternVL3-78B & 50.5 & \color{teal}{+16.4} & 46.2 & \color{red}{-6.7} & 44.3 & \color{red}{-1.4} & 47.0 \\
        & Qwen3.5-397B-A17B & 37.4 & \color{red}{-14.6} & 64.3 & \color{teal}{+1.8} & 42.4 & \color{red}{-13.2} & 48.0 \\
        & Qwen3.5-122B-A10B & 36.1 & \color{red}{-12.9} & 62.5 & \color{teal}{+1.3} & 35.2 & \color{red}{-16.7} & 44.6 \\
        & Qwen3.5-35B-A3B & 33.4 & \color{red}{-12.3} & 64.4 & \color{teal}{+6.6} & 33.9 & \color{red}{-16.7} & 43.9 \\
        & Qwen3.5-9B & 32.8 & \color{red}{-6.7} & 63.0 & \color{teal}{+5.9} & 33.9 & \color{red}{-18.4} & 43.2 \\
        & Kimi-VL-A3B & 34.9 & \color{teal}{+12.8} & 54.3 & \color{teal}{+15.4} & 32.6 & \color{red}{-4.8} & 40.6 \\
        & MiniCPM-o-4.5 & 19.3 & \color{red}{-10.5} & 50.5 & \color{teal}{+2.6} & 29.2 & \color{red}{-18.9} & 33.0 \\
        \bottomrule
    \end{tabular}
\end{table}
A summary of results for in-context learning is given in Table~\ref{tab:icl}.
\paragraph{Finding 3: In-context demonstrations improve frontier models and Gemma.}
Since zero-shot models lack domain-specific knowledge (Finding~1), in-context demonstrations can inject this knowledge directly by providing labeled examples of each scenario type.
All seven closed-source models improve on every axis with in-context demonstrations.
Claude Opus 4.6 gains +28.9 points on Q6 (60.5$\to$89.4), and Gemini-3.1-Pro gains +18.6 (71.2$\to$89.8).
Notably, Gemma-4-31B-IT is the \emph{only open model} we benchmarked that benefits comparably, gaining +23.9 on Q6 (62.1$\to$86.0), matching closed-model behavior while being open-weight.
The improvements are largest on Q6 (calibration diagnosis), the most practically relevant task, where labeled examples provide the status vocabulary the model needs.

\paragraph{Finding 4: Multi-image demonstrations degrade open-weight models.}
In contrast, Qwen3.5 models, MiniCPM-o, and Kimi-VL are \emph{actively harmed} by multi-image demonstrations: Qwen3.5-9B drops 18.4~points on Q6, and Qwen3.5-35B-A3B drops 16.7.
The degradation is specific to $N$-way multi-image prompts: 1-shot parameter extraction (Q5), which uses a single demonstration, improves for nearly all models.
This suggests these model families can process a single reference image but fail to relate multiple labeled demonstrations to a query. 

\paragraph{Finding 5: $N$-way scaling shows the MM-ICL gap is not a simple image-overload effect.}
To determine whether degradation under in-context learning is simply caused by image overload, we conduct $N$-way scaling experiments (0--5 shots, 1,008 QA pairs; details in Appendix~\ref{app:nway}).
Frontier models improve consistently with more demonstrations (Gemini: 44$\to$85; Gemma-4-31B: 49$\to$82), proving the task supports multi-image reasoning.
In contrast, Qwen3.5, MiniCPM-o, and Kimi-VL all peak at 1-shot then degrade, and larger Qwen models do not recover this behavior (397B peaks at 35 vs.\ 9B at 44).
These results show that the MM-ICL gap is not simply a consequence of too many images or task difficulty: reducing the number of demonstrations does not eliminate the failure pattern in these model families.

\paragraph{Failure patterns.}
Qualitative analysis (Appendix~\ref{app:failure-patterns}) reveals recurring failure modes: (1)~\emph{visual similarity confusion}, where models correctly describe plots but misclassify scenarios differing in subtle features like oscillation frequency; (2)~\emph{optimistic bias}, where 85\% of Q6 errors predict success when ground truth indicates failure; (3)~\emph{``No fit'' blindness}, where models report ``Reliable'' for raw data without fitted curves, affecting 38\% of samples.

\section{SFT Ablation and Case Study}
\label{sec:method}

\subsection{Ablation Study: What Can Fine-Tuning Improve?}

For the ablation study using Qwen3.5 9B models, we use a \emph{train/test-family split}: models are trained on one set of experiment families and evaluated on previously unseen families, rather than on a standard sample-level partition.
We generate two SFT datasets by applying QCalEval's question templates to synthetic calibration plots from the train families, paired with ground-truth answers: \textbf{Zero-shot} (25.8K zero-shot QA pairs) and \textbf{ICL} (12.9K ICL QA pairs), and evaluate five recipes: zero-shot-only, ICL-only, blend, ICL$\rightarrow$zero-shot, and zero-shot$\rightarrow$ICL.
Full results are in Appendix~\ref{app:sft}.

\paragraph{Key findings from the ablation study.}
We find that: (1)~\textbf{SFT substantially improves zero-shot scores}: zero-shot SFT improves Q6 from 61.1 to 70.6 (+9.5 points), while sequential curricula raise Q4 from 28.6 to about 60 in the train/test-family ablation.
(2)~\textbf{In-context learning remains difficult after SFT}: different recipes improve different classification axes, but gains are modest and inconsistent across Q5 and Q6.
(3)~\textbf{No recipe improves Q3 under in-context learning}: the best result (24.1) remains below the base score (27.1), suggesting that free-text scientific reasoning may require advances beyond supervised fine-tuning, such as reasoning-oriented training or reinforcement learning from expert feedback.
(4)~\textbf{Sequential training order matters}: among sequential curricula, ICL$\rightarrow$zero-shot is the strongest overall in the train/test-family ablation. We hypothesize that training on zero-shot data first teaches the model to rely solely on the query image; subsequent ICL training then fails to override this habit, as the model has already learned it can answer without referencing demonstration images.

\subsection{NVIDIA Ising Calibration 1}
\label{sec:model}

Guided by the ablation study, we release \textbf{NVIDIA Ising Calibration 1}, an open-weight Qwen3.5-35B-A3B (MoE) model~\cite{qwen3.5} trained with the strongest sequential SFT curriculum identified in that study (ICL$\rightarrow$zero-shot).
Model configuration details are in Appendix~\ref{app:casestudy}; weights are at \url{https://huggingface.co/nvidia/Ising-Calibration-1-35B-A3B}.

\begin{table}[t]
    \centering
    \caption{NVIDIA Ising Calibration 1 scores on QCalEval compared to its base model. \textbf{Bold} = best per column.}
    \label{tab:ising}
    \small
    \begin{tabular}{lccccccc|ccc}
        \toprule
        & \multicolumn{7}{c|}{\textbf{Zero-Shot}} & \multicolumn{3}{c}{\textbf{ICL}} \\
        \cmidrule(lr){2-8} \cmidrule(lr){9-11}
        \textbf{Model} & \textbf{Q1} & \textbf{Q2} & \textbf{Q3} & \textbf{Q4} & \textbf{Q5} & \textbf{Q6} & \textbf{Avg} & \textbf{Q3} & \textbf{Q5} & \textbf{Q6} \\
        \midrule
        Qwen3.5-35B base & 86.8 & 39.9 & 45.7 & 52.7 & 57.8 & 50.6 & 55.5 & \textbf{33.4} & \textbf{64.4} & 33.9 \\
        \textbf{Ising-Cal-1} & \textbf{87.8} & \textbf{67.1} & \textbf{64.7} & \textbf{90.5} & \textbf{62.5} & \textbf{75.3} & \textbf{74.7} & 31.2 & 59.8 & \textbf{42.4} \\
        \midrule
        $\Delta$ & \color{teal}{+1.0} & \color{teal}{+27.2} & \color{teal}{+19.0} & \color{teal}{+37.8} & \color{teal}{+4.7} & \color{teal}{+24.7} & \color{teal}{+19.2} & \color{red}{-2.2} & \color{red}{-4.6} & \color{teal}{+8.5} \\
        \bottomrule
    \end{tabular}
\end{table}

Table~\ref{tab:ising} shows that the benchmark-derived SFT recipe substantially improves the base Qwen3.5-35B model in zero-shot evaluation, with the clearest gains on diagnosis, scientific interpretation, and fit-related judgments.
The improvement is broad but not uniform: zero-shot plot understanding benefits strongly from domain tuning, whereas in-context learning remains weak and largely retains the same multi-demonstration failure pattern as the base model.
This makes Ising Calibration 1 a stronger reference model for single-plot understanding on QCalEval, but not a solution to the MM-ICL gap.

\section{Limitations}
\label{sec:limitations}

\paragraph{Dataset size and coverage.}
QCalEval contains 243 samples across 22 experiment families, mainly covering superconducting qubits and neutral atoms, with 186 simulated and 57 hardware samples.
Future work can expand the benchmark to cover more experiment types, hardware platforms, and modalities.
The per-family sample sizes (2--21 samples) limit the statistical power to draw fine-grained per-experiment conclusions.

\section{Conclusion}
\label{sec:conclusion}

We introduced QCalEval, the first comprehensive benchmark for VLMs on quantum calibration plots, evaluating 18 models across six question types under both zero-shot and in-context learning settings.
Our zero-shot evaluation reveals a two-step capability gap: models detect visual features well (Q1: 65--91\%) but lack the domain knowledge to map them to operational outcomes (Q2: 32--67\%, Q6: 37--75\%), with systematic optimistic bias toward predicting success. Fit assessment (Q4) reveals a distinct visual judgment gap, in which weaker models cannot distinguish reliable from unreliable fits.
In-context learning, which injects domain knowledge through labeled visual examples, improves frontier models, but many open-weight models degrade under multi-image prompts.
A systematic SFT ablation at the 9-billion-parameter scale shows that supervision format is critical: sequential curricula raise Q4 from 28.6 to about 60, and Q6 from 61.1 to 70.6 in zero-shot evaluation, but zero-shot SFT degrades in-context learning performance, and no configuration closes the gap on free-text scientific reasoning (Q3).
As a case study, we release NVIDIA Ising Calibration 1, an open model based on Qwen3.5-35B-A3B (MoE) and trained using a two-phase sequential SFT recipe.
Reliable calibration plot understanding is a prerequisite for autonomous quantum computing workflows; by releasing QCalEval and Ising Calibration 1 as open resources, we enable the community to benchmark progress and fine-tune models for their specific hardware platforms and experiment types.

\section*{Data and Code Availability}
The QCalEval benchmark dataset is available at \url{https://huggingface.co/datasets/nvidia/QCalEval}.
Evaluation scripts are available at \url{https://github.com/nvidia/QCalEval}.
NVIDIA Ising Calibration 1 model weights are available at \url{https://huggingface.co/nvidia/Ising-Calibration-1-35B-A3B}.

\section*{Acknowledgments}
We thank Shi Xuan Leong for contributions to early discussions, and Lilian Zhong for assistance with data verification.
This manuscript has been authored by Fermi Research Alliance, LLC under Contract No.\ DE-AC02-07CH11359 with the U.S.\ Department of Energy, Office of Science, Office of High Energy Physics. G. H., N. V. and Y. X. acknowledged the support from the U.S. Department of Energy, Office of Science, National Quantum Information Science Research Centers, Quantum Systems Accelerator (Award No. DESCL0000121), and Advanced Scientific Computing Research Testbeds for Science program under Contract No.
DE-AC02-05CH11231. A. G. F. and A. V. acknowledged the support from Business Finland through project CfoQ (787/31/2025).
A.A. and I.R. acknowledge the support of the UK government Department for Science, Innovation and Technology through the UK National Quantum Technologies Programme, and of the Engineering and Physical Sciences Research Council (Grant No. EP/Z53318X/1: QCI3 Hub). L.M.C. acknowledges funding from the Novo Nordisk Foundation, Grant number NNF22SA0081175, NNF Quantum Computing Programme. A.A.G. thanks Anders G. Frøseth for his generous support and acknowledges the generous support of Natural Resources Canada and the Canada 150 Research Chairs program. A.A.-G. and V.B. acknowledge the University of Toronto’s Acceleration Consortium, which receives funding from the CFREF-2022-00042 Canada First Research Excellence Fund. A.A.-G. acknowledges support from the AI2050 program from the Schmidt Foundation. D.C.C. gratefully acknowledges contributions from Infleqtion team members Garrett Hickman, Kevin Kuper, David Mason, and Peter Mitchell.
\bibliographystyle{unsrtnat}
\bibliography{references}

\definecolor{nvgreen}{HTML}{76B900}
\definecolor{nvgreendark}{HTML}{4A7A00}
\definecolor{nvgreenbg}{HTML}{F4F9EC}
\newcommand{\cmark}{\textcolor{nvgreen}{\ding{51}}}
\newcommand{\xmark}{\textcolor{gray}{\ding{55}}}
\newcommand{\pmark}{\textcolor{gray}{\ding{115}}}
\newcommand{\qsep}{\par\vspace{3pt}\noindent\textcolor{gray}{\rule{\linewidth}{0.5pt}}\par\vspace{3pt}}
\newcommand{\qheader}[3]{\par\vspace{2pt}\noindent\textbf{#1} \textbf{#2} \hfill \textbf{#3}\par\vspace{4pt}}
\newcommand{\qlabel}[1]{\makebox[4.5em][l]{\textbf{#1}}}
\newcommand{\qnote}{\makebox[4.5em][l]{\textbf{Analysis:}}}
\appendix

\section{Dataset Details}
\label{app:dataset}

\subsection{Experiment Family Provenance}
\label{app:families}

QCalEval spans 22 experiment families, primarily covering superconducting qubits and neutral atoms.
Table~\ref{tab:app-families} provides the complete breakdown by group, experiment family, and sample count.

\begin{table}[h]
    \centering
    \caption{Experiment families in QCalEval by group and sample count.}
    \label{tab:app-families}
    \small
    \begin{tabular}{llr}
        \toprule
        \textbf{Group} & \textbf{Experiment Family} & \textbf{Samples} \\
        \midrule
        \multirow{15}{*}{Superconducting}
        & Resonator Spectroscopy & 15 \\
        & Qubit Spectroscopy & 9 \\
        & Qubit Spectroscopy (2D Power-Freq) & 18 \\
        & Qubit Flux Spectroscopy & 18 \\
        & Coupler Flux Spectroscopy & 4 \\
        & Rabi Oscillation (Simulated) & 21 \\
        & Rabi Oscillation (Hardware) & 11 \\
        & DRAG Calibration & 15 \\
        & T1 Relaxation & 12 \\
        & T1 Fluctuations & 9 \\
        & Ramsey (Frequency Cal) & 12 \\
        & Ramsey (T2*) & 15 \\
        & Ramsey Charge Tomography & 12 \\
        & PingPong Calibration & 12 \\
        & Single-Shot Readout & 15 \\
        \midrule
        \multirow{6}{*}{Neutral Atom}
        & MOT Loading & 9 \\
        & Tweezer Array & 2 \\
        & Rydberg Spectroscopy & 2 \\
        & Rydberg Ramsey & 5 \\
        & Microwave Ramsey & 6 \\
        & CZ Benchmarking & 6 \\
        \midrule
        Electron-on-Helium & Pinch-off & 15 \\
        \midrule
        \multicolumn{2}{l}{\textbf{Total}} & \textbf{243} \\
        \bottomrule
    \end{tabular}
\end{table}

\paragraph{Leakage prevention.}
For the ablation study, the \emph{train/test-family split} is performed at the experiment-family level rather than the individual-sample level: all samples from a family appear in either train or test, never both, so test families are genuinely unseen during training.
This prevents the model from training on one calibration family and then being evaluated on visually near-identical variants of that same family.
For NVIDIA Ising Calibration 1, training covers all experiment families but uses only synthetic training images; no benchmark images appear in the training set.
In-context learning benchmark construction uses a different safeguard: support examples are always distinct samples from the query itself, and singleton scenario types are excluded so the target example is never reused as its own demonstration.

\subsection{Benchmark Statistics}
\label{app:stats}

The benchmark contains 243 samples spanning 87 scenario types, with 309 unique images (some samples contain multiple images).
Each sample includes six question-answer pairs (Q1--Q6), yielding 1,458 total QA pairs for single-turn evaluation.

\paragraph{Q6 label distribution.}
The Q6 calibration diagnosis labels vary by experiment family, with each family defining two to seven scenario types.
Most families have one success scenario and multiple failure modes (e.g., DRAG has one success and four distinct failure types), so the benchmark is weighted toward failure cases, reflecting the practical priority of failure detection in calibration workflows.

\paragraph{In-context learning exclusions.}
Seven scenario types with only one sample are excluded from in-context learning evaluation because no within-family support examples exist.
This reduces the in-context learning evaluation set to 236 samples.

\subsection{Annotation Workflow}
\label{app:annotation}

Ground-truth answers were created through a human-AI collaborative process:

\begin{enumerate}
    \item \textbf{Expert seeding}: Domain experts provided brief scenario descriptions and key observations for each calibration plot. These per-sample annotations served as the shared ground truth from which all six question-answer pairs were derived.

    \item \textbf{AI expansion}: Both GPT-5.4 and Gemini-3.1-Pro independently expanded the expert annotations into complete Q1--Q6 answers, generating structured JSON for Q1/Q5, explanatory text for Q3, and status labels for Q2/Q4/Q6.

    \item \textbf{Cross-validation}: The two model outputs were compared to identify discrepancies. Cases where the models disagreed were flagged for human review.

    \item \textbf{Key-point generation}: For Q1 and Q3 scoring, key-point checklists were generated by both models and cross-validated to create the rubrics used by the GPT-5.4 judge.

    \item \textbf{Human verification}: Experts reviewed all flagged cases and a random sample of agreeing cases, correcting errors in both answers and key-point checklists to ensure scientific accuracy.
\end{enumerate}

\section{Evaluation Protocol Details}
\label{app:protocol}

\subsection{Inference Settings}

\paragraph{Inference.}
In zero-shot evaluation, each of the six questions is sent as an independent single-turn request with the plot image; no conversation history is shared between questions for the same sample.
In in-context learning evaluation, each QA pair (Q3, Q5, Q6) is likewise an independent conversation with demonstration images and labels prepended.
All models are queried with greedy decoding (temperature~$= 0$) and a maximum output budget of 16{,}384 tokens.
We use uniform decoding settings across all models for fair comparison, even though some vendors recommend model-specific hyperparameters for multimodal tasks (e.g., Qwen3.5 suggests \texttt{top\_p\,=\,0.001} and \texttt{repetition\_penalty\,=\,1.05} for vision inputs).
Vendor-tuned settings may improve individual model scores but would confound cross-model comparisons.
Closed-source models are accessed via API; open-weight models are served with vLLM~\cite{kwon2023vllm}.
The exact zero-shot and in-context learning prompt templates are released with the benchmark; the appendix summarizes the prompt structure and includes representative template excerpts, while the released benchmark files remain the authoritative source.
For reproducibility, we also report the exact public model identifiers and access date for the closed models evaluated in this paper.
The API-based closed-model evaluations reported here use the versions available on April 6, 2026: \texttt{gpt-5.4}, \texttt{gpt-5.4-mini}, \texttt{gemini-3.1-pro-preview}, \texttt{gemini-3.1-flash-lite-preview}, \texttt{claude-opus-4-6}, and \texttt{claude-sonnet-4-6}.

\subsection{Scoring Methods}
\label{app:scoring}

\paragraph{Overview.}
Four of six axes use fully deterministic scoring with no judge model (Q2, Q4, Q5, Q6).
Two axes require free-text evaluation (Q1, Q3) and are scored by two independent LLM judges---GPT-5.4 and Gemini 3.1 Pro Preview---with the final score averaged across both judges to reduce single-judge bias.
Programmatic scores (Q2, Q4, Q5, Q6) are identical across judges.

Each question type uses a specific scoring method:

\paragraph{Q1: Visual Perception (Figure Description).}
Score = 50\% programmatic + 50\% LLM key-point matching (averaged across GPT-5.4 and Gemini 3.1 Pro Preview judges).
The programmatic component checks exact match on \texttt{plot\_type}, \texttt{x\_axis.scale}, and \texttt{y\_axis.scale}.
The GPT component evaluates whether the response captures 3--5 key visual elements (e.g., ``decay visible,'' ``two clusters separated'') from a human-verified checklist.

\paragraph{Q2: Domain Comprehension (Outcome Classification).}
4-way exact match accuracy.
Labels: \texttt{Expected behavior}, \texttt{Suboptimal parameters}, \texttt{Anomalous behavior}, \texttt{Apparatus issue}.
The predicted label must exactly match the ground truth after case normalization.

\paragraph{Q3: Scientific Reasoning (Significance Analysis).}
Each of the two LLM judges (GPT-5.4 and Gemini 3.1 Pro Preview) independently scores whether the response addresses 3 key points from a human-verified checklist.
Each key point is scored 0/0.5/1 (missing/partial/full), and the final Q3 score is the average across both judges, yielding a 0--100 score.
The judge prompt instructs evaluation of scientific content, not writing style.

\paragraph{Q4: Fit Assessment (Validity Check).}
3-way exact match accuracy.
Labels: \texttt{Reliable}, \texttt{Unreliable}, \texttt{No fit}.
Evaluates whether the model correctly assesses fit quality from visual inspection.

\paragraph{Q5: Quantitative Extraction (Parameter Extraction).}
Per-field tolerance scoring on JSON output.
Each field is scored based on type-specific tolerances:
\begin{itemize}
    \item \texttt{pct}: Percentage tolerance
    \item \texttt{enum}: Exact categorical match
    \item \texttt{bool}: Boolean match
    \item \texttt{int\_count} / \texttt{count\_float}: count-valued fields with full-credit and half-credit tolerances
    \item \texttt{abs}: absolute-value tolerance
    \item \texttt{coord\_list}: coordinate lists scored element-wise with tolerance
    \item \texttt{array\_int\_match}: variable-length integer arrays scored by tolerance-aware F1
    \item \texttt{array\_float\_match}: variable-length float arrays scored by tolerance-aware matching
\end{itemize}
Failed JSON parses score 0. The final score averages across all fields.

\paragraph{Q6: Calibration Diagnosis (Status Classification).}
Multi-way exact match accuracy.
Each experiment family defines 2--7 possible status labels (e.g., \texttt{SUCCESS}, \texttt{NO\_SIGNAL}, \texttt{AMP\_TOO\_HIGH}).
Labels are normalized by stripping whitespace, converting to uppercase, and mapping common synonyms.

\subsection{Prompt Templates}
\label{app:prompts}

\paragraph{Prompt sources.}
The exact prompt strings used in zero-shot and in-context learning evaluation are stored verbatim in the released benchmark files.
The evaluation scripts do not prepend a hidden system prompt: each benchmark call is a single user message with image blocks interleaved with the released text.

\paragraph{Zero-shot prompts.}
Each Q1--Q6 request is an independent user turn.
Q1 uses the exact JSON-description template below:

\begin{tcolorbox}[colback=white, colframe=OliveGreen, colbacktitle=OliveGreen, coltitle=white, title=Zero-Shot Q1 Template (verbatim structure), fonttitle=\bfseries\small, breakable, boxrule=0.8pt, left=4pt, right=4pt]
\small\ttfamily
Describe the figure <image> in JSON format.

Required fields:
\{
  "plot\_type": "scatter" | "line" | "heatmap" | "histogram",
  "x\_axis": \{"label": string, "scale": "linear" | "log", "range": [min, max]\},
  "y\_axis": \{"label": string, "scale": "linear" | "log", "range": [min, max]\},
  "main\_features": string
\}
\end{tcolorbox}

For Q2--Q6, the prompt prepends the experiment-family background and then asks the task-specific question about the query image(s).

\paragraph{In-context learning prompts.}
In-context learning evaluation also uses a single user message with interleaved support images and text.
The three prompt families for in-context learning evaluation are:
\begin{itemize}
    \item \textbf{Q3 analysis:} background + one example analysis per scenario type in the family + query image(s).
    \item \textbf{Q5 extraction:} background + one sibling example from the same scenario type + query image(s).
    \item \textbf{Q6 classification:} background + one labeled example per scenario type + query image(s).
\end{itemize}

A representative exact structure for Q6 in-context learning classification is:
\begin{tcolorbox}[colback=white, colframe=OliveGreen, colbacktitle=OliveGreen, coltitle=white, title=In-Context Learning Q6 Template (generator structure), fonttitle=\bfseries\small, breakable, boxrule=0.8pt, left=4pt, right=4pt]
\small\ttfamily
\{background\}

Classify this chart given the following labeled examples:

<image> Status: \{status\_1\}
Suggested range: \{range\_1\}

<image> Status: \{status\_2\}
Suggested range: \{range\_2\}

\ldots

Now classify this chart:
<image>
\end{tcolorbox}

For Q3 and Q6, the number of support examples is one representative example per available scenario type in the family (excluding the query sample itself when possible); Q5 always uses exactly one sibling example from the same scenario type.

\paragraph{N-way classification prompt.}
The $N$-way benchmark uses the following template pattern:

\begin{tcolorbox}[colback=white, colframe=OliveGreen, colbacktitle=OliveGreen, coltitle=white, title=N-Way Q6 Template (generator structure), fonttitle=\bfseries\small, breakable, boxrule=0.8pt, left=4pt, right=4pt]
\small\ttfamily
\{background\}

This experiment has the following possible statuses:
- \{STATUS\_1\}
- \{STATUS\_2\}
- \ldots

[optional labeled support examples]

Now classify this chart:
<image>
\end{tcolorbox}

\subsection{Decoding and Transport Settings}
\label{app:inference}

All models were evaluated with the following settings unless otherwise noted:
\begin{itemize}
    \item \textbf{Temperature}: 0.0 (deterministic decoding)
    \item \textbf{Max tokens}: 16,384
    \item \textbf{Retries}: Up to 3 retries on API failures
    \item \textbf{Top-p}: not explicitly overridden in the released scripts; provider defaults apply
    \item \textbf{Image transport}: PNG images encoded as base64 data URLs, converted to RGBA when required for API compatibility
\end{itemize}

For multi-image prompts under in-context learning evaluation, images are interleaved with text in chronological order (support examples first, query last).

\section{Model Details}
\label{app:models}

Table~\ref{tab:app-models} lists the reported models with their specifications.

\begin{table}[h]
    \centering
    \caption{Model specifications for reported models.}
    \label{tab:app-models}
    \small
    \begin{tabular}{llll}
        \toprule
        \textbf{Model} & \textbf{Provider} & \textbf{Parameters} & \textbf{Type} \\
        \midrule
        GPT-5.4 & OpenAI & --- & Closed \\
        GPT-5.4-Mini & OpenAI & --- & Closed \\
        Gemini-3.1-Pro & Google & --- & Closed \\
        Gemini-3.1-Flash-Lite & Google & --- & Closed \\
        Claude Opus 4.6 & Anthropic & --- & Closed \\
        Claude Sonnet 4.6 & Anthropic & --- & Closed \\
        Claude Haiku 4.5 & Anthropic & --- & Closed \\
        \midrule
        Qwen3.5-397B-A17B & Alibaba & 397B MoE & Open \\
        Qwen3.5-122B-A10B & Alibaba & 122B MoE & Open \\
        Qwen3.5-35B-A3B & Alibaba & 35B MoE & Open \\
        Qwen3.5-27B & Alibaba & 27B & Open \\
        Qwen3.5-9B & Alibaba & 9B & Open \\
        Gemma-4-31B-IT & Google & 31B & Open \\
        InternVL3-78B & Shanghai AI Lab & 78B & Open \\
        InternVL3-38B & Shanghai AI Lab & 38B & Open \\
        MiniCPM-o-4.5 & OpenBMB & 9B & Open \\
        Kimi-VL-A3B & Moonshot & 16B MoE & Open \\
        \midrule
        Ising-Cal-1 & NVIDIA & 35B MoE & Open (domain-tuned) \\
        \bottomrule
    \end{tabular}
\end{table}

\section{Extended Results}
\label{app:results}

\subsection{Full Per-Family Breakdown}
\label{app:perfamily}

Table~\ref{tab:app-perfamily-full} provides the complete Q6 classification score breakdown across 22 experiment families for four representative models: two closed-source (Gemini-3.1-Pro, GPT-5.4) and two open-weight (Gemma-4-31B-IT, Qwen3.5-35B-A3B).

\begin{table*}[t]
    \centering
    \caption{Q6 classification scores by experiment family. Families are sorted by Gemini-3.1-Pro score. $n$ = number of benchmark samples in the family.}
    \label{tab:app-perfamily-full}
    \scriptsize
    \begin{tabular}{p{0.28\textwidth}rcccc}
        \toprule
        \textbf{Experiment Family} & $n$ & \textbf{Gemini-3.1} & \textbf{GPT-5.4} & \textbf{Gemma-4-31B} & \textbf{Qwen3.5-35B} \\
        \midrule
        Coupler Flux Spectroscopy & 4 & 100.0 & 75.0 & 100.0 & 75.0 \\
        Resonator Spectroscopy & 15 & 100.0 & 100.0 & 100.0 & 100.0 \\
        T1 Fluctuations & 9 & 100.0 & 100.0 & 77.8 & 88.9 \\
        MOT Loading & 9 & 100.0 & 88.9 & 88.9 & 77.8 \\
        Tweezer Array & 2 & 100.0 & 100.0 & 50.0 & 100.0 \\
        Rydberg Ramsey & 5 & 100.0 & 80.0 & 60.0 & 80.0 \\
        Microwave Ramsey (Neutral Atom) & 6 & 100.0 & 83.3 & 83.3 & 50.0 \\
        Rabi Oscillation (Hardware) & 11 & 90.9 & 81.8 & 63.6 & 36.4 \\
        Qubit Spectroscopy & 9 & 88.9 & 55.6 & 55.6 & 44.4 \\
        CZ Benchmarking & 6 & 83.3 & 83.3 & 66.7 & 66.7 \\
        T1 Relaxation & 12 & 75.0 & 50.0 & 50.0 & 50.0 \\
        Ramsey Charge Tomography & 12 & 75.0 & 75.0 & 75.0 & 66.7 \\
        Pinch-off & 15 & 73.3 & 40.0 & 73.3 & 66.7 \\
        DRAG Calibration & 15 & 66.7 & 73.3 & 80.0 & 20.0 \\
        Rabi Oscillation (Simulated) & 21 & 66.7 & 66.7 & 71.4 & 52.4 \\
        Qubit Spectroscopy (2D Power-Freq) & 18 & 66.7 & 61.1 & 72.2 & 44.4 \\
        Qubit Flux Spectroscopy & 18 & 61.1 & 55.6 & 50.0 & 38.9 \\
        Single-Shot Readout & 15 & 53.3 & 20.0 & 26.7 & 20.0 \\
        Ramsey (Frequency Cal) & 12 & 50.0 & 25.0 & 25.0 & 25.0 \\
        Rydberg Spectroscopy & 2 & 50.0 & 50.0 & 50.0 & 50.0 \\
        PingPong Calibration & 12 & 33.3 & 58.3 & 50.0 & 41.7 \\
        Ramsey (T2*) & 15 & 33.3 & 20.0 & 20.0 & 26.7 \\
        \bottomrule
    \end{tabular}
\end{table*}

\subsection{Confusion Analysis}
\label{app:confusion}

\paragraph{Q2 outcome classification.}
Models show a systematic optimistic bias on Q2 (4-way outcome classification).
When the ground truth is not \texttt{Expected behavior}, errors frequently collapse into \texttt{Expected behavior} or \texttt{Suboptimal parameters} rather than the more severe labels \texttt{Anomalous behavior} and \texttt{Apparatus issue}.
The confusion is asymmetric: models are much less likely to flip clear successes into severe failure categories.

\paragraph{Q6 diagnosis classification.}
The same optimistic bias appears in Q6.
Models tend to overpredict \texttt{SUCCESS} on ambiguous plots, especially for readout, Ramsey, and spectroscopy families where failure modes differ by subtle geometric or contrast cues.
This pattern is consistent across model families and scales, suggesting that current VLMs have difficulty distinguishing subtle failure signatures from genuinely successful calibrations.

\subsection{Universally Misclassified Failure Modes}
\label{app:q2-universal}

On Q2 (4-way outcome classification), \textbf{24 of 87 scenario types} score below 10\% across \emph{all} base models---meaning every model misclassifies these samples, typically as ``Expected behavior.''
These universally misclassified scenario types are exclusively failure modes, listed in Table~\ref{tab:app-q2-universal}.

\begin{table}[h]
    \centering
    \caption{Scenarios where all base models score $<$10\% on Q2 outcome classification. All are failure modes misclassified as ``Expected behavior.'' Grouped by experiment family.}
    \label{tab:app-q2-universal}
    \scriptsize
    \begin{tabular}{ll}
        \toprule
        \textbf{Family} & \textbf{Universally Misclassified Scenarios} \\
        \midrule
        Coupler Flux & \texttt{failure\_bad\_fit} \\
        CZ Benchmarking & \texttt{failure\_miscalibrated} \\
        DRAG & \texttt{failure\_no\_signal} \\
        Single-Shot Readout & \texttt{failure\_no\_excitation}, \texttt{failure\_no\_signal} \\
        Microwave Ramsey & \texttt{failure\_detuned}, \texttt{failure\_low\_contrast} \\
        MOT Loading & \texttt{tailed} \\
        Pinchoff & \texttt{failure\_noisy\_no\_transition}, \texttt{failure\_stabilize\_positive} \\
        Qubit Flux Spec & \texttt{failure\_not\_tunable} \\
        Rabi & \texttt{failure\_too\_fast} \\
        Ramsey Freq Cal & \texttt{beating}, \texttt{too\_few\_osc}, \texttt{too\_many\_osc} \\
        Ramsey T2* & \texttt{beating}, \texttt{window\_too\_short} \\
        Res Spec & \texttt{wide\_scan\_no\_signal}, \texttt{zoomed\_no\_signal} \\
        Rydberg Spec & \texttt{failure\_low\_contrast} \\
        T1 & \texttt{failure\_window\_too\_short} \\
        Tweezer Array & \texttt{failure\_aberrated} \\
        \bottomrule
    \end{tabular}
\end{table}

This is not a model-specific weakness but a systematic domain gap: no current VLM can reliably distinguish these calibration failures from expected behavior using a coarse 4-way taxonomy.

\section{Qualitative Analysis: Universally Misclassified Failure Modes}
\label{app:failure-patterns}

As shown in Table~\ref{tab:app-q2-universal}, 24 of 87 scenario types are misclassified by \emph{all} base models on Q2.
We present four representative cases in depth, showing all six question--answer pairs to illustrate \emph{where} in the reasoning pipeline each model fails.
Each case uses real prompts, images, and model responses from a single benchmark evaluation run.
Even at temperature 0 (greedy decoding), API-based models may not produce identical outputs across runs due to non-deterministic inference backends.
The programmatic scores (Q2, Q4, Q5, Q6) are stable across the majority of repeated runs, but individual responses may vary in wording.

\subsection{Case Study 1: DRAG Calibration --- No Signal}

\textbf{Entry:} \texttt{drag\_failure\_no\_signal\_a} \quad \qlabel{Model:} GPT-5.4 (score: 32.2/100)

The DRAG sweep shows broadly scattered data with no consistent linear trend. Both fitted lines are nearly flat---there is no physically meaningful zero-crossing. GPT-5.4 scores 32.2/100, failing on Q2, Q3, Q5, and Q6.

\begin{tcolorbox}[colback=white, colframe=OliveGreen, colbacktitle=OliveGreen, coltitle=white, title=DRAG No-Signal: GPT-5.4 (32.2/100), fonttitle=\bfseries\small, breakable, boxrule=1pt, left=6pt, right=6pt, top=6pt, bottom=6pt]
\small
\begin{center}
\includegraphics[width=0.55\textwidth]{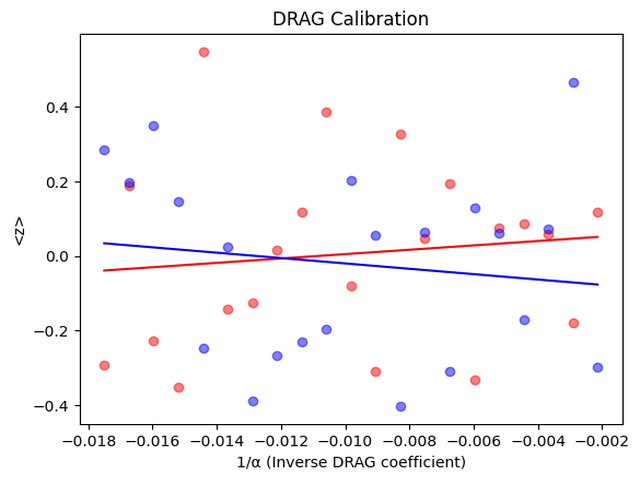}
\end{center}

\textbf{Background (shared across Q2--Q6):} ``This is a DRAG calibration: we sweep 1/alpha to find the optimal value that minimizes leakage. A successful result has the zero-crossing of fitted curves clearly observable in the sweep window.''
\qsep
\qheader{Q1}{Visual Description}{85/100}
\qlabel{Prompt:} ``Describe the figure in JSON format. Required fields: plot\_type, x\_axis, y\_axis, main\_features.''

\qlabel{Model:} Correctly identifies scatter plot with two datasets and linear fits.
\qsep
\qheader{Q2}{Outcome Classification}{0/100}
\qlabel{Prompt:} ``Classify the experimental outcome. Options: Expected behavior / Suboptimal parameters / Anomalous behavior / Apparatus issue.''

\qlabel{GT:} Apparatus issue --- complete lack of signal response.

\qlabel{Model:} ``\texttt{Suboptimal parameters} --- The fitted red and blue trends do show opposite slopes and a zero-crossing, but the crossing is weak and not clearly resolved against the large scatter.''

\noindent\qnote Recognizes poor signal quality but underestimates severity to ``needs adjustment'' rather than apparatus failure.
\qsep
\qheader{Q3}{Scientific Analysis}{8/100}
\qlabel{Prompt:} ``What does this result imply? Explain: what the zero-crossing position and slope indicate; whether the sweep range captures the crossing; what calibration step follows.''

\qlabel{Model:} ``This result looks inconclusive / weakly calibrated... the blue fit has a small negative slope, the red has a small positive slope...''

\noindent\qnote Hedges but still treats the crossing as physically meaningful.
\qsep
\qheader{Q4}{Fit Reliability}{100/100}
\qlabel{Prompt:} ``Assess whether the fit is reliable for parameter extraction. Options: Reliable / Unreliable / No fit.''

\qlabel{Model:} ``\texttt{Unreliable} --- The fitted lines have weak slopes and the zero-crossing is not clearly constrained.''
\qsep
\qheader{Q5}{Parameter Extraction}{0/100}
\qlabel{Prompt:} ``Extract parameters in JSON: \{optimal\_alpha\_inv: float, intersection\_clear: bool\}.''

\qlabel{GT:} \texttt{\{``optimal\_alpha\_inv'': ``Unreliable'', ``intersection\_clear'': false\}}

\qlabel{Model:} \texttt{\{``optimal\_alpha\_inv'': -0.0117, ``intersection\_clear'': true\}}

\noindent\qnote Extracts a specific number from noise.
\qsep
\qheader{Q6}{Calibration Diagnosis}{0/100}
\qlabel{Prompt:} ``Determine experiment status. Criteria: SUCCESS / NO\_SIGNAL / OPTIMAL\_NOT\_CENTERED.''

\qlabel{GT:} \texttt{NO\_SIGNAL}

\qlabel{Model:} ``\texttt{SUCCESS} --- The fitted curves clearly cross within the sweep window, around $1/\alpha \approx -0.012$.''

\noindent\qnote The ``crossing'' is an artifact of fitting random scatter.
\end{tcolorbox}

\textbf{Pattern:} The model consistently over-interprets noise. It correctly notes weak fits (Q4) but still extracts parameters and declares success, showing a disconnect between fit assessment and downstream reasoning.

\subsection{Case Study 2: Single-Shot Readout --- No Excitation}

\textbf{Entry:} \texttt{gmm\_failure\_no\_excitation\_a} \quad \qlabel{Model:} GPT-5.4 (score: 61.1/100)

The IQ scatter plot shows two overlapping clusters with nearly identical population distributions in both panels (80/20 vs 76/24)---the excitation drive failed to change the qubit state. GPT-5.4 scores 61.1/100, passing Q1/Q3/Q4/Q5 but failing Q2 and Q6.

\begin{tcolorbox}[colback=white, colframe=OliveGreen, colbacktitle=OliveGreen, coltitle=white, title=Single-Shot Readout No-Excitation: GPT-5.4 (61.1/100), fonttitle=\bfseries\small, breakable, boxrule=1pt, left=6pt, right=6pt, top=6pt, bottom=6pt]
\small
\begin{center}
\includegraphics[width=0.55\textwidth]{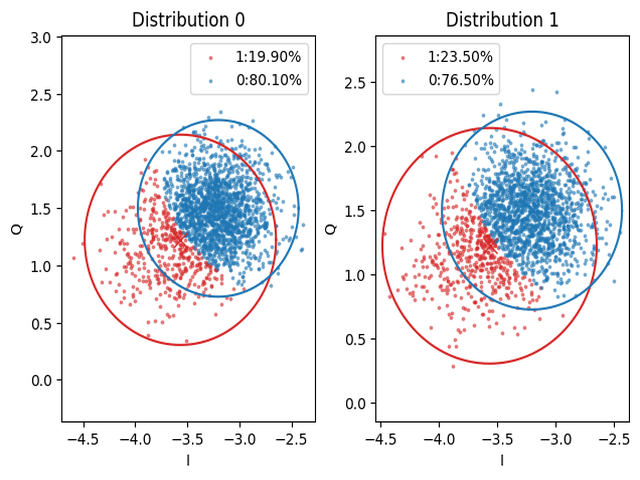}
\end{center}

\textbf{Background (shared across Q2--Q6):} ``This is a single-shot readout discrimination experiment: the I-Q scatter plot shows measurement results for $|0\rangle$ and $|1\rangle$ states fitted with a Gaussian Mixture Model. A successful result has two well-separated clusters.''
\qsep
\qheader{Q1}{Visual Description}{100/100}
\qlabel{Prompt:} ``Describe the figure in JSON format.''

\qlabel{Model:} Correctly describes two scatter plots with overlapping clusters.
\qsep
\qheader{Q2}{Outcome Classification}{0/100}
\qlabel{Prompt:} ``Classify the experimental outcome. Options: Expected behavior / Suboptimal parameters / Anomalous behavior / Apparatus issue.''

\qlabel{GT:} Anomalous behavior --- excitation drive not effective.

\qlabel{Model:} ``\texttt{Suboptimal parameters} --- The readout is functioning, but clusters overlap substantially.''

\noindent\qnote Sees overlap as a readout problem, misses that excitation itself failed.
\qsep
\qheader{Q3}{Scientific Analysis}{92/100}
\qlabel{Prompt:} ``What does this result imply? Explain: cluster separation and overlap; whether discrimination is sufficient; what adjustment follows.''

\qlabel{Model:} ``This result looks marginal to poor for single-shot discrimination... clusters are not well separated...''

\noindent\qnote Mostly correct reasoning, though doesn't identify the no-excitation root cause.
\qsep
\qheader{Q4}{Fit Reliability}{100/100}
\qlabel{Prompt:} ``Assess whether the fit is reliable. Options: Reliable / Unreliable / No fit.''

\qlabel{Model:} ``\texttt{Unreliable} --- The two Gaussian clusters overlap substantially.''
\qsep
\qheader{Q5}{Parameter Extraction}{75/100}
\qlabel{Prompt:} ``Extract parameters in JSON: \{separation, cluster0\_center, cluster1\_center\}.''

\qlabel{Model:} \texttt{\{``separation'': ``overlapping'', ``cluster0\_center'': [-3.64, 1.23], ``cluster1\_center'': [-3.17, 1.50]\}}

\noindent\qnote Separation correct; cluster centers partially within tolerance.
\qsep
\qheader{Q6}{Calibration Diagnosis}{0/100}
\qlabel{Prompt:} ``Determine experiment status. Criteria: SUCCESS / NO\_SIGNAL / NO\_EXCITATION / HIGH\_POWER / NO\_RES\_RESPONSE.''

\qlabel{GT:} \texttt{NO\_EXCITATION} --- both distributions have same populations.

\qlabel{Model:} ``\texttt{SUCCESS} --- Both panels show two distinct clusters with visibly separated centers.''

\noindent\qnote Sees ``two clusters'' and declares success without comparing population ratios across panels.
\end{tcolorbox}

\textbf{Pattern:} The model has strong visual perception (Q1/Q5) and reasonable analysis (Q3) but cannot make the domain-specific inference that identical population ratios across preparations means no excitation occurred.

\subsection{Case Study 3: Ramsey Frequency Calibration --- Beating}

\textbf{Entry:} \texttt{ramsey\_failure\_freq\_cal\_beating\_a} \quad \qlabel{Model:} Gemini-3.1-Pro (score: 40.3/100)

The Ramsey fringes show amplitude modulation (beating) from two closely-spaced frequency components, visible as waxing/waning contrast across the time window. The FFT shows a dominant peak but the beating makes the single-frequency fit unreliable. Gemini-3.1-Pro scores 40.3/100.

\begin{tcolorbox}[colback=white, colframe=OliveGreen, colbacktitle=OliveGreen, coltitle=white, title=Ramsey Beating: Gemini-3.1-Pro (40.3/100), fonttitle=\bfseries\small, breakable, boxrule=1pt, left=6pt, right=6pt, top=6pt, bottom=6pt]
\small
\begin{center}
\begin{minipage}{0.48\textwidth}
\centering
\includegraphics[width=\textwidth]{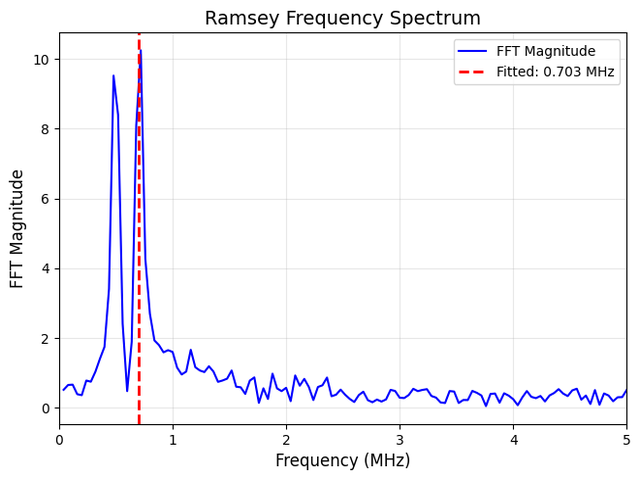}\\
\small FFT spectrum
\end{minipage}
\hfill
\begin{minipage}{0.48\textwidth}
\centering
\includegraphics[width=\textwidth]{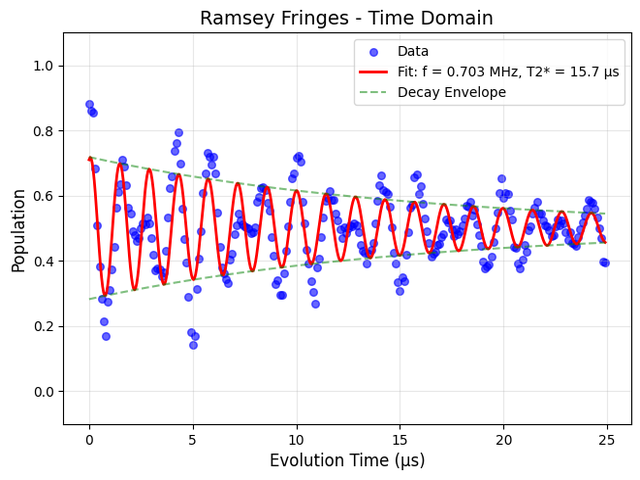}\\
\small Time-domain fringes
\end{minipage}
\end{center}

\textbf{Background (shared across Q2--Q6):} ``This is a Ramsey frequency calibration experiment: two $\pi$/2 pulses separated by a variable delay measure frequency detuning. A successful result shows clear oscillations at the detuning frequency with a fit that accurately extracts the frequency offset.''
\qsep
\qheader{Q1}{Visual Description}{100/100}
\qlabel{Prompt:} ``Describe each figure in JSON format (2 images).''

\qlabel{Model:} Correctly describes both plots (FFT + time domain).
\qsep
\qheader{Q2}{Outcome Classification}{0/100}
\qlabel{Prompt:} ``Classify the experimental outcome. Options: Expected behavior / Suboptimal parameters / Anomalous behavior / Apparatus issue.''

\qlabel{GT:} Anomalous behavior --- beating indicates multiple frequency components.

\qlabel{Model:} ``\texttt{Expected behavior} --- Clear, well-defined Ramsey fringes successfully fitted, extracting a distinct detuning frequency.''

\noindent\qnote Sees ``clear fringes'' and ignores the amplitude modulation envelope.
\qsep
\qheader{Q3}{Scientific Analysis}{8/100}
\qlabel{Prompt:} ``What do these results imply? Explain: what the oscillation pattern indicates about detuning and coherence; whether the measurement is sufficient; what parameter adjustment follows.''

\qlabel{Model:} ``The data shows clear oscillations... the frequency spectrum confirms a strong, unambiguous peak at the fitted frequency.''

\noindent\qnote Completely misses the beating signature despite it being visible in the time-domain plot.
\qsep
\qheader{Q4}{Fit Reliability}{100/100}
\qlabel{Prompt:} ``Assess whether the fit is reliable. Options: Reliable / Unreliable / No fit.''

\qlabel{Model:} ``\texttt{Unreliable} --- The fit does not capture the primary oscillation frequency present in the data. The data shows a lower frequency oscillation that is completely missed by the fit.''

\noindent\qnote Correct conclusion with partially correct reasoning (identifies fit/data mismatch, though the root cause is beating).
\qsep
\qheader{Q5}{Parameter Extraction}{33/100}
\qlabel{Prompt:} ``Extract parameters in JSON: \{T2\_star\_us, detuning\_MHz, fringes\_visible\}.''

\qlabel{GT:} \texttt{\{``T2\_star\_us'': ``Unreliable'', ``detuning\_MHz'': ``Unreliable'', ``fringes\_visible'': 18\}}

\qlabel{Model:} \texttt{\{``T2\_star\_us'': 15.7, ``detuning\_MHz'': 0.703, ``fringes\_visible'': 17\}}

\noindent\qnote Reads fitted values literally instead of recognizing they are unreliable due to beating.
\qsep
\qheader{Q6}{Calibration Diagnosis}{0/100}
\qlabel{Prompt:} ``Determine experiment status. Criteria: SUCCESS / NO\_DETUNING / BEATING / TOO\_MANY\_OSC / TOO\_FEW\_OSC.''

\qlabel{GT:} \texttt{BEATING} --- amplitude modulation from multiple frequencies.

\qlabel{Model:} ``\texttt{SUCCESS} --- Clear Ramsey fringes with a well-defined decay envelope.''

\noindent\qnote Does not recognize that envelope modulation indicates beating, not simple decay.
\end{tcolorbox}

\textbf{Pattern:} The model reads fitted parameter values from the plot annotation but cannot independently assess whether those values are physically trustworthy. Beating---a common calibration failure---is invisible to models that equate ``oscillations + fit'' with success.

\subsection{Case Study 4: Coupler Flux Spectroscopy --- Bad Fit}

\textbf{Entry:} \texttt{coupler\_flux\_failure\_bad\_fit\_a} \quad \qlabel{Model:} GPT-5.4 (score: 33.3/100)

Two side-by-side heatmaps show a tunable coupler's frequency dispersion vs.\ bias voltage. The raw spectroscopy data is clean---bright branches with resolved avoided crossings are clearly visible. However, the overlaid fit curves systematically miss the measured branches, placing sharp asymptotes at wrong positions. GPT-5.4 scores 33.3/100, failing on Q2, Q3, Q4, and Q6.

\begin{tcolorbox}[colback=white, colframe=OliveGreen, colbacktitle=OliveGreen, coltitle=white, title=Coupler Flux Bad Fit: GPT-5.4 (33.3/100), fonttitle=\bfseries\small, breakable, boxrule=1pt, left=6pt, right=6pt, top=6pt, bottom=6pt]
\small
\begin{center}
\includegraphics[width=0.7\textwidth]{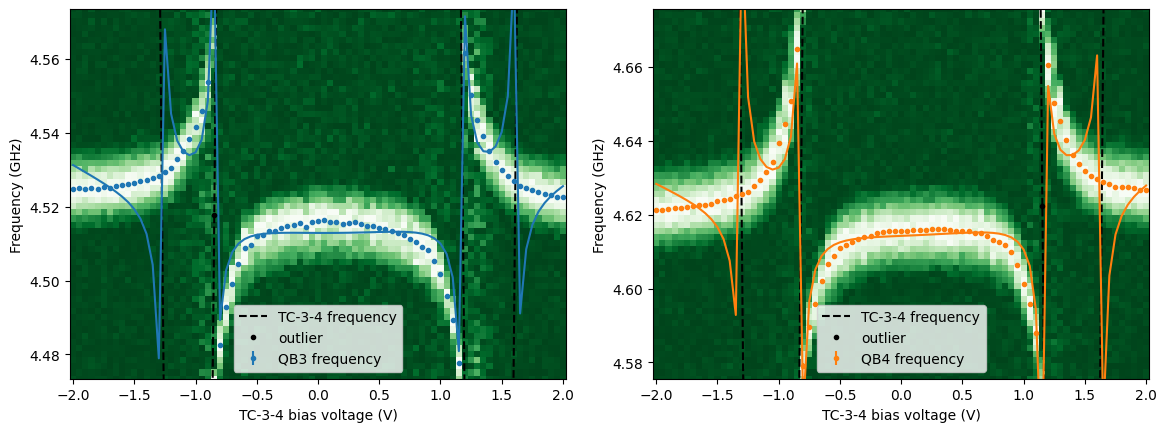}
\end{center}

\textbf{Background (shared across Q2--Q6):} ``This is tunable coupler spectroscopy: we map the coupler's frequency response vs applied flux bias. A successful result shows a clear coupler dispersion curve with a good fit.''
\qsep
\qheader{Q1}{Visual Description}{100/100}
\qlabel{Prompt:} ``Describe the figure in JSON format.''

\qlabel{Model:} Correctly identifies two heatmaps with bias voltage and frequency axes, bright curved bands, and overlaid fits.
\qsep
\qheader{Q2}{Outcome Classification}{0/100}
\qlabel{Prompt:} ``Classify the experimental outcome. Options: Expected behavior / Suboptimal parameters / Anomalous behavior / Apparatus issue.''

\qlabel{GT:} Suboptimal parameters --- the fit model has failed to capture the dispersion.

\qlabel{Model:} ``\texttt{Expected behavior} --- Both panels show a clear, continuous tunable-coupler dispersion versus bias with strong spectroscopy contrast.''

\noindent\qnote Declares success because the \emph{data} is clean, ignoring that the \emph{fit} is wrong.
\qsep
\qheader{Q3}{Scientific Analysis}{0/100}
\qlabel{Prompt:} ``What does this result imply? Explain: what the dispersion curve reveals about tunability and coupling strength; whether the ranges are sufficient; what next step follows.''

\qlabel{Model:} ``This looks like a successful tunable-coupler spectroscopy measurement... the coupler is alive and tunable...''

\noindent\qnote Completely misses the fit failure, treating good data as good calibration.
\qsep
\qheader{Q4}{Fit Reliability}{0/100}
\qlabel{Prompt:} ``Assess whether the fit is reliable. Options: Reliable / Unreliable / No fit.''

\qlabel{GT:} \texttt{Unreliable} --- fit curves fail to trace the avoided crossings.

\qlabel{Model:} ``\texttt{Reliable} --- The fitted curves closely follow the bright spectroscopy features.''

\noindent\qnote Contradicts visual evidence: fit visibly deviates at avoided crossings.
\qsep
\qheader{Q5}{Parameter Extraction}{100/100}
\qlabel{Prompt:} ``Extract parameters in JSON: \{crossing\_voltages\_V, left\_fig\_branch\_freqs\_GHz, right\_fig\_branch\_freqs\_GHz\}.''

\qlabel{GT:} \texttt{\{``crossing\_voltages\_V'': [-0.85, 1.15], ``left\_fig\_branch\_freqs\_GHz'': [4.52, 4.52, 4.52], ``right\_fig\_branch\_freqs\_GHz'': [4.62, 4.62, 4.63]\}}

\qlabel{Model:} \texttt{\{``crossing\_voltages\_V'': [-0.86, 1.15], ``left\_fig\_branch\_freqs\_GHz'': [4.525, 4.516, 4.523], ``right\_fig\_branch\_freqs\_GHz'': [4.622, 4.616, 4.628]\}}

\noindent\qnote All fields within scoring tolerances.
\qsep
\qheader{Q6}{Calibration Diagnosis}{0/100}
\qlabel{Prompt:} ``Determine experiment status. Criteria: SUCCESS / FIT\_POOR.''

\qlabel{GT:} \texttt{FIT\_POOR} --- dispersion visible but fit deviates systematically.

\qlabel{Model:} ``\texttt{SUCCESS} --- The fitted curves closely follow the measured bright spectroscopy features with only minor local deviations.''

\noindent\qnote Cannot distinguish ``data looks good'' from ``fit is accurate.''
\end{tcolorbox}

\textbf{Pattern:} This case illustrates the model's inability to critically evaluate fit quality against raw data. When both the data and a fit are present, the model defaults to ``the fit follows the data'' without checking whether the fit actually captures the key features (avoided crossings, branch structure). This is particularly dangerous because the data quality is high---the failure is entirely in the fitting, not the measurement.

\section{$N$-Way Classification Scaling}
\label{app:nway}

To understand how models leverage increasing numbers of demonstrations, we evaluate Q6 classification as labeled examples scale from 0 to 5 shots on four experiment families (DRAG, Rabi, Ramsey, T1) with 5 scenario types each, yielding 1,008 total QA pairs.

\begin{table}[h]
    \centering
    \caption{$N$-way Q6 classification scores as the number of labeled demonstrations increases (0--5 shot). 4~experiment families with 5~scenario types each, 1{,}008~QA pairs total.}
    \label{tab:app-nway}
    \small
    \begin{tabular}{llcccccc}
        \toprule
        & \textbf{Model} & \textbf{0} & \textbf{1} & \textbf{2} & \textbf{3} & \textbf{4} & \textbf{5} \\
        \midrule
        \multirow{7}{*}{\rotatebox{90}{\scriptsize Closed}}
        & Claude Opus 4.6 & \textbf{55.6} & \textbf{85.7} & 77.8 & 76.2 & 78.3 & 82.5 \\
        & Gemini-3.1-Pro & 44.4 & 82.0 & 79.4 & 77.8 & \textbf{85.2} & \textbf{84.7} \\
        & Claude Sonnet 4.6 & 47.6 & 67.7 & 70.4 & 69.3 & 76.7 & 70.9 \\
        & GPT-5.4 & 44.4 & 58.2 & 64.0 & \textbf{76.7} & 75.7 & 77.8 \\
        & Gemini-3.1-Flash-Lite & 42.9 & 50.8 & 54.0 & 65.6 & 64.6 & 67.2 \\
        & GPT-5.4-Mini & 34.9 & 47.1 & 52.4 & 47.6 & 54.0 & 52.9 \\
        & Claude Haiku 4.5 & 38.1 & 42.9 & 48.7 & 43.9 & 53.4 & 68.8 \\
        \midrule
        \multirow{10}{*}{\rotatebox{90}{\scriptsize Open}}
        & Gemma-4-31B-IT & 49.2 & 69.3 & 72.5 & 77.2 & 80.4 & 81.5 \\
        & MiniCPM-o-4.5 & 25.4 & 72.5 & 32.8 & 27.5 & 19.6 & 18.5 \\
        & Kimi-VL-A3B & 20.6 & 64.6 & 46.0 & 29.6 & 26.5 & 20.6 \\
        & InternVL3-78B & 27.0 & 57.7 & 32.8 & 37.0 & 38.6 & 42.3 \\
        & Qwen3.5-27B & 25.4 & 47.1 & 25.9 & 23.3 & 25.4 & 27.0 \\
        & Qwen3.5-35B-A3B & 23.8 & 46.6 & 30.7 & 27.0 & 21.2 & 19.6 \\
        & Qwen3.5-9B & 31.7 & 44.4 & 35.4 & 26.5 & 22.8 & 22.8 \\
        & Qwen3.5-122B-A10B & 22.2 & 37.0 & 37.0 & 23.8 & 23.3 & 21.2 \\
        & Qwen3.5-397B-A17B & 33.3 & 34.9 & 28.0 & 17.5 & 23.8 & 21.7 \\
        & InternVL3-38B & 17.5 & 31.2 & 30.7 & 32.8 & 40.7 & 39.7 \\
        \bottomrule
    \end{tabular}
\end{table}

\paragraph{Two distinct learning behaviors.}
Table~\ref{tab:app-nway} shows two patterns across 17 models.
Frontier closed models and Gemma-4-31B improve consistently with more demonstrations: Gemini-3.1-Pro rises from 44.4 (0-shot) to 84.7 (5-shot), and Gemma-4-31B from 49.2 to 81.5.
Qwen3.5-based models peak at 1-shot and then degrade with additional examples --- this pattern holds across all five Qwen3.5 variants tested (9B, 27B, 35B-A3B, 122B-A10B, 397B-A17B), with the largest 397B model showing the weakest 1-shot peak (34.9) and steeper degradation.
MiniCPM-o-4.5 and Kimi-VL-A3B show similar peak-then-degrade behavior, suggesting this pattern extends beyond the Qwen family.
Reducing the number of demonstrations does not remove the degradation for these model families, which is inconsistent with a simple image-count explanation.

\section{SFT Ablation Details}
\label{app:sft}

This section provides complete details on the supervised fine-tuning experiments summarized in the main text.

\subsection{Training Data}
\label{app:sft-data}
We generate two complementary SFT datasets from the 9 train experiment families using a shared data generation pipeline.

\paragraph{Zero-shot data.}
The zero-shot training data is generated via a three-step pipeline:
\begin{enumerate}
    \item \textbf{Synthetic plot generation}: For each experiment family, we study the real partner data and build mathematical simulators that replicate the same plot format and physical behavior across all scenario types, with known ground-truth parameters (e.g., frequencies, decay rates, fit quality).
    \item \textbf{Metadata pairing}: Each synthetic plot is paired with its scenario label and the ground-truth parameters used to generate it.
    \item \textbf{LLM augmentation}: Qwen3.5-397B-A17B rewrites the structured metadata into natural expert-style answers for all six question types.
\end{enumerate}
This produces 25.8K zero-shot QA pairs for the ablation study.

\paragraph{ICL-formatted data.}
The ICL training data is constructed by sampling from the synthetic images generated above and formatting them as multi-image demonstrations following the benchmark ICL prompt structure. For Q3 (scientific analysis) and Q6 (calibration diagnosis), $N$-way demonstrations are prepended with one labeled example per scenario type in the experiment family. For Q5 (parameter extraction), a 1-shot example from a sibling of the same scenario type is provided. Ground-truth answers serve as demonstration labels. This yields 12.9K ICL-formatted QA pairs for the ablation study.

\subsection{SFT Recipes}
We evaluate five data configurations:
(1)~\textbf{Zero-shot}: SFT on zero-shot QA pairs only;
(2)~\textbf{ICL}: SFT on ICL QA pairs only;
(3)~\textbf{Blend}: zero-shot and ICL QA pairs are merged into a single mixed training set and optimized jointly in one training run;
(4)~\textbf{ICL$\rightarrow$Zero-shot}: sequential training, ICL first then zero-shot;
(5)~\textbf{Zero-shot$\rightarrow$ICL}: sequential training, zero-shot first then ICL.
Here, \emph{Blend} denotes joint training on a mixed pool of zero-shot and ICL-formatted examples, whereas the sequential recipes expose the model to the two formats in separate stages.
We evaluate four learning rates ($10^{-6}$, $2{\times}10^{-6}$, $5{\times}10^{-6}$, $10^{-5}$) and report the best-performing configuration for each recipe under each evaluation mode.

\subsection{Zero-Shot SFT Results}

\begin{table}[h]
    \centering
    \caption{SFT ablation: zero-shot scores under the train/test-family split (Qwen3.5-9B).}
    \label{tab:app-sft-zeroshot}
    \small
    \begin{tabular}{lccccccc}
        \toprule
        \textbf{Recipe} & \textbf{Q1} & \textbf{Q2} & \textbf{Q3} & \textbf{Q4} & \textbf{Q5} & \textbf{Q6} & \textbf{Avg} \\
        \midrule
        Base & 82.0 & 42.9 & 41.1 & 28.6 & 66.1 & 61.1 & 53.6 \\
        Zero-shot & 82.1 & 47.6 & 44.2 & 55.6 & 69.3 & \textbf{70.6} & 61.6 \\
        ICL & 81.8 & 53.2 & 42.3 & 28.6 & \textbf{70.3} & 57.1 & 55.6 \\
        Blend & \textbf{83.6} & 54.0 & 44.6 & 34.9 & 69.2 & 62.7 & 58.2 \\
        ICL$\rightarrow$ZS & 80.5 & 52.4 & 46.3 & 58.7 & 69.5 & 65.9 & \textbf{62.2} \\
        ZS$\rightarrow$ICL & 83.2 & \textbf{56.3} & \textbf{46.6} & \textbf{60.3} & 62.0 & 60.3 & 61.4 \\
        \bottomrule
    \end{tabular}
\end{table}

Table~\ref{tab:app-sft-zeroshot} shows that domain-specific SFT yields substantial zero-shot gains on previously unseen experiment families.
The strongest single-format recipe is zero-shot-only SFT, which improves Q6 classification from 61.1 to 70.6 and reaches 61.6 average.
Sequential curricula further improve fit assessment and overall transfer, raising Q4 from 28.6 to about 60 and producing the strongest overall train/test-family result with ICL$\rightarrow$zero-shot at 62.2 average (+8.6 over base).
These results indicate that recipe order matters, and that ICL$\rightarrow$zero-shot is the strongest sequential curriculum in the 9B train/test-family ablation.

\subsection{ICL SFT Results}

\begin{table}[h]
    \centering
    \caption{SFT ablation: in-context learning scores under the train/test-family split (Qwen3.5-9B).}
    \label{tab:app-sft-icl}
    \small
    \begin{tabular}{lcccc}
        \toprule
        \textbf{Scale} & \textbf{Recipe} & \textbf{Q3} & \textbf{Q5} & \textbf{Q6} \\
        \midrule
        \multicolumn{5}{l}{\textit{Qwen3.5-9B}} \\
        9B & Base & \textbf{27.1} & 76.0 & 37.3 \\
        9B & Zero-shot & 23.3 & 76.7 & \textbf{42.1} \\
        9B & ICL & 23.9 & \textbf{84.3} & 32.5 \\
        9B & Blend & 17.6 & 70.2 & 38.9 \\
        9B & ICL$\rightarrow$ZS & 24.1 & 77.4 & 38.1 \\
        9B & ZS$\rightarrow$ICL & 16.3 & 67.8 & 36.5 \\
        \bottomrule
    \end{tabular}
\end{table}

Table~\ref{tab:app-sft-icl} reveals that in-context learning remains difficult after SFT.
Single-format SFT improves specific classification axes: ICL-formatted SFT gives the best Q5 score (84.3), while zero-shot SFT gives the best Q6 score (42.1).
However, \textbf{no SFT recipe improves Q3 under in-context learning}: the best result (24.1 with ICL$\rightarrow$ZS) remains below the base score (27.1).
Overall in-context learning gains are modest and inconsistent across recipes.

\section{Case Study: Model and Training Details}
\label{app:casestudy}

NVIDIA Ising Calibration 1 is based on Qwen3.5-35B-A3B, a mixture-of-experts (MoE) VLM with 35B total parameters but only 3B active per token. We apply a two-phase sequential supervised fine-tuning (SFT) recipe guided by the 9B ablation study described in Appendix~\ref{app:sft}. The following subsections describe the training data generation process and the training recipe.

\subsection{Training Data Generation}
\label{app:datagen}

The training data is generated using the same pipeline described in Appendix~\ref{app:sft-data}, but covering all 22 experiment families rather than the 9 train families used in the ablation study. This produces 48.7K zero-shot QA pairs and 23.8K ICL-formatted QA pairs.

\subsection{Training Recipe}
\label{app:training-recipe}

\paragraph{Phase 1: ICL-formatted SFT.}
The model is first trained on the 23.8K ICL-formatted QA pairs (lr $= 10^{-5}$, 1 epoch), teaching it to process multi-image demonstrations and relate labeled examples to a query plot.

\paragraph{Phase 2: Zero-shot SFT.}
Training continues on the 48.7K LLM-augmented zero-shot QA pairs (lr $= 5 \times 10^{-6}$, 1 epoch). The lower learning rate preserves the ICL capabilities acquired in Phase~1 while strengthening single-plot understanding.

\paragraph{Optimization.}
Both phases use AdamW ($\beta_1{=}0.9$, $\beta_2{=}0.999$, $\epsilon{=}10^{-8}$, weight decay $= 0$) with cosine learning-rate decay and 3\% linear warmup. Training is full-parameter SFT with a frozen vision tower, BF16 precision, and an effective batch size of 128.

\paragraph{Availability.}
Open weights are available at \url{https://huggingface.co/nvidia/Ising-Calibration-1-35B-A3B}.

\end{document}